\documentclass{article}

\usepackage{arxiv}

\usepackage[utf8]{inputenc} % allow utf-8 input
\usepackage[T1]{fontenc}    % use 8-bit T1 fonts
\usepackage{hyperref}       % hyperlinks
\usepackage{url}            % simple URL typesetting
\usepackage{booktabs}       % professional-quality tables
\usepackage{amsfonts}       % blackboard math symbols
\usepackage{amsmath}
\usepackage{nicefrac}       % compact symbols for 1/2, etc.
\usepackage{microtype}      % microtypography
\usepackage{xcolor}
\usepackage{tabularx}
\usepackage{algorithm}
\usepackage[noend]{algpseudocode}
\usepackage{listings}
\usepackage{cleveref}       % smart cross-referencing (load after the packages it refers into)

\usepackage{graphicx}
\usepackage{tikz}            % draft diagrams (fig:pipeline)
\usetikzlibrary{arrows.meta,positioning,fit,calc}
\usepackage{natbib}
\usepackage{doi}

\title{ICAD: Search-based Algorithm for CAD Reconstruction}

\usepackage{authblk}

\newbox{\orcid}\sbox{\orcid}{\includegraphics[scale=0.06]{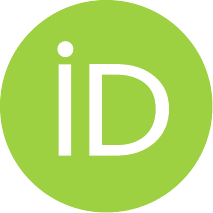}}
\newcommand{\orcidauthor}[2]{\mbox{\href{https://orcid.org/#1}{\usebox{\orcid}\hspace{1mm}#2}}}
\author[1]{\orcidauthor{0000-0003-3802-1774}{Albert Garifullin}\thanks{Corresponding author: \texttt{albert.garifullin@graphics.cs.msu.ru}}}
\author[2]{\orcidauthor{0009-0001-2677-881X}{Nikolay Mayorov}}
\author[1]{Nikolay Petrov}% TODO: ORCID, then \orcidauthor{<id>}
\author[1]{Dmitrii Zhemchuzhnikov} % TODO: ORCID, then \orcidauthor{<id>}
\author[1]{Soslan Kabisov} % TODO: ORCID, then \orcidauthor{<id>}
\author[1,2]{\orcidauthor{0000-0001-8829-9884}{Vladimir Frolov}}
\affil[1]{Lomonosov Moscow State University, 119991 Moscow, Russia}
\affil[2]{Institute for Artificial Intelligence of Lomonosov Moscow State University (IAI MSU), 119192 Moscow, Russia}

\renewcommand{\shorttitle}{ICAD: Search-based Algorithm for CAD Reconstruction}
\date{}

\hypersetup{
pdftitle={ICAD: Search-based Algorithm for CAD Reconstruction},
pdfsubject={cs.CG, cs.CV},
pdfauthor={Albert Garifullin, Nikolay Mayorov, Nikolay Petrov, Dmitrii Zhemchuzhnikov, Soslan Kabisov, Vladimir Frolov},
pdfkeywords={CAD reverse engineering, parametric modeling, CSG},
}

\begin{document}
\maketitle

\begin{abstract}
We propose a search-based approach to CAD reverse engineering that recovers a CAD program from a triangle mesh.  Our method outperform state-of-the-art accuracy being generally faster than existing approaches, and running on a standard desktop CPU. The pipeline is multi-stage: First, primitive detectors extract candidate extrusions and revolutions from the target mesh to seed an initial population; a genetic algorithm then searches the program space by mutating the discrete structure, while in each generation, gradient descent refines the continuous parameters of the top half of the population using analytic gradients from a differentiable signed distance field (SDF) evaluator; finally, a post-processing pass prunes negligible features, simplifies profiles, and fits arcs. Our pipeline don't rely on any pre-trained data, although it can use such data inside primitive detector module, and this give additional accuracy when the data is available. On the output our optimizer map CAD model to CadQuery code and takes the form of a construction history: a sequence of features, where each feature extrudes or revolves a 2D sketch (composed of line and arc loops), places it in 3D space, and applies a Boolean union, difference, or intersection.

\end{abstract}


\section{Introduction}
\label{sec:intro}

A parametric computer-aided design (CAD) model is effectively a program that represents a tree of Boolean operations whose leaves are extruded and revolved sketches. Because this constructive representation is typically proprietary to the originating CAD system, parts are instead exchanged across systems as meshes or boundary representations (B-Rep). Consequently, many legacy models have lost their procedural definitions over time. CAD reverse engineering aims to recover these underlying programs from unstructured observations such as images, point clouds, or meshes.

Despite significant recent efforts in CAD reverse engineering, the problem remains open. Single-pass networks \citep{kabisov2026cadena} rely on scaling up training rather than optimizing directly against the target geometry, which limits their precision. Conversely, methods that execute and score candidates achieve higher accuracy only at substantial computational cost \citep{savrasov2026cadfather,benchcad2026leaderboard}, while still relying heavily on non-learned search or geometric heuristics~\citep{nehme2026stepcad,savrasov2026cadfather}. Pure search without learning, meanwhile, has long been considered too slow and narrow for complex parts \citep{fayolle2016evolutionary,du2018inversecsg}.

We address the CAD reverse engineering problem in the mesh-to-CAD setting, where the input is a triangle mesh of a part and the output is a CAD program whose solid matches it. 
Our approach is a search-based engine that operates without any pre-trained models. 
We achieve state-of-the-art results by combining a hypothesis generator (via geometric detectors), genetic search in the space of \textbf{CAD programs}, 
and gradient-descent optimization of the continuous program parameters using a differentiable Signed Distance Function (SDF) representation. 
The resulting CAD program maps to CadQuery code. Our contribution is:
\begin{itemize}
	\item SDF-based kernel for differentiable CAD programs, which map closely to CadQuery;
	\item hypothesis generation, which detects different primitives in the target mesh and combines them into complete programs that seed the search;
	\item an evolutionary search over programs that changes their structure with genetic operators and refines their continuous parameters by gradient descent through a differentiable CSG evaluator;
\end{itemize}

\section{Related Work}
\label{sec:related}

Mesh-to-CAD reconstruction seeks an executable program whose result matches a target shape. We first place the
problem within inverse procedural modeling, then review three lines of work: detection-based methods, which infer the
program in a single pass; search-based methods, which look for a program that explains the target; and iterative and
agentic methods, which execute partial programs and revise them using feedback. We end with methods that optimize the
continuous parameters of a given program.

\paragraph{Inverse procedural modeling.}
Recovering a procedure that generates a given shape is known as inverse procedural modeling. \citet{stava2010lsystems}
infer L-systems from vector drawings, while Metropolis procedural modeling~\citep{talton2011metropolis} and
stochastically-ordered sequential Monte Carlo~\citep{ritchie2015sosmc} search the derivations of a procedural model
for one that meets a high-level specification. \citet{du2018inversecsg} distinguish two strategies: recognition, which
infers structure directly from the input and is fast but gives limited insight into it, and search, which explains the
input fully at a higher cost per shape. Learning can support either. \citet{ellis2018graphics} detect drawing
primitives with a network and use program synthesis to find a program with loops and conditionals that draws them, and
\citet{ellis2019repl} guide a search over CSG programs with a policy and a value function that observe the execution of
partial programs. \citet{ritchie2023neurosymbolic} survey such neurosymbolic models. CAD reconstruction is the case
where the procedure is a CAD program and the specification is a target solid.

\paragraph{Detection-based methods.}
These methods follow the recognition strategy: they infer the program in a single pass, by deterministic rules or by a
trained network. Classical conversion from a boundary representation to CSG~\citep{shapiro1991csg} writes the solid
as a union of canonical intersection terms over the halfspaces bounding its faces, and \citet{buchele2004halfspace}
extend it to curved faces given in implicit form. These deductive methods need an exact B-rep, and their trees are
rarely compact~\citep{du2018inversecsg}. On meshes and point clouds, primitives are detected first, for example with
efficient RANSAC~\citep{schnabel2007ransac}, which finds planes, spheres, cylinders, cones and tori. Learned detectors
output primitive assemblies directly: CSGNet~\citep{sharma2018csgnet} decodes a CSG program with a recurrent network;
CAPRI-Net~\citep{yu2022capri} intersects quadric primitives into convex parts and combines these by union and
difference; Point2Cyl~\citep{uy2022point2cyl} segments a point cloud into extrusion cylinders and fits their
parameters in closed form; and SECAD-Net~\citep{li2023secad} learns implicit sketches and extrusion parameters by
self-supervision and unites the extruded solids. Their outputs are structured, but they are not in general CAD code with an
explicit modeling sequence.

Most recent learned methods predict that sequence directly. DeepCAD~\citep{wu2021deepcad} defines a command language
for sketch-and-extrude models, learns a Transformer autoencoder over 178k of them, and maps point clouds into its latent
space. CAD-Recode~\citep{rukhovich2025cadrecode} attaches a point-cloud projector to a pretrained language model that
writes CadQuery code and trains it on a million procedurally generated programs; cadrille~\citep{kolodiazhnyi2025cadrille}
extends this approach to images and text and fine-tunes it with reinforcement learning on geometric rewards.
CAD-Coder~\citep{doris2025cadcoder} generates CadQuery code from images with a vision-language model, and
CADEvolve~\citep{elistratov2026cadevolve} grows a training corpus of 1.3M programs by evolutionary synthesis. These
models condition on the target rather than optimize against it, so the precision of their continuous parameters is
limited by what the network has learned. Our method also starts with detection: extrusions and revolutions detected on
the mesh seed its search, as hypotheses to be combined and refined rather than as the answer. It needs no training
data; the evolution in CADEvolve produces data offline, whereas ours runs at test time on a single target.

\paragraph{Search-based methods.}
InverseCSG~\citep{du2018inversecsg} made search practical on meshes by separating the continuous part of the problem
from the discrete one. It detects an over-complete set of solid primitives (spheres, cylinders, cuboids and tori),
which fixes all continuous parameters; places a labeled sample in each region cut out by the primitives' surfaces; and
lets a SAT-based synthesizer find the smallest tree consistent with the samples, within each segment of a
divide-and-conquer decomposition. A post-process simplifies the tree and re-parameterizes it for editing, and the
method handles models with over a hundred primitives. \citet{wu2018csg} instead label voxels as inside or outside a
point cloud and build the tree bottom-up by energy minimization. Evolutionary methods search the space of trees
stochastically. After early genetic algorithms for CSG~\citep{hamza2004csg,weiss2009thesis},
\citet{fayolle2016evolutionary} evolve set-theoretic expressions over primitives fitted to a point cloud, with a
fitness measured at the points and a penalty on tree size. \citet{friedrich2019gecco} speed this up with a
deterministic filter of the points, a partition of the problem into independent subproblems, and redundancy removal
followed by local search, and later combine such steps with several metaheuristics in a pipeline that also optimizes
the editability of the tree~\citep{friedrich2020pipeline}. \citet{fayolle2024survey} survey this field. InverseCSG
argued against evolutionary search: in its comparison, the genetic algorithm of \citet{fayolle2016evolutionary} needed
at least two hours per iteration on parts with 19--42 primitives and was stopped after 200 hours, and its
non-determinism made it hard to control. Our method belongs to this line. Like InverseCSG, it separates detection from
search, and like the evolutionary methods, it searches with a genetic algorithm. It differs from both in three ways:
its vocabulary is extrusions and revolutions of sketches, which map one-to-one to CadQuery, rather than solid
primitives; its search also moves the continuous parameters instead of fixing them at detection; and its fitness is
volumetric, whereas that of \citet{fayolle2016evolutionary} is defined on the surface.

\paragraph{Iterative and agentic methods.}
A third line executes partial programs and uses the result to decide what to do next. The Fusion~360
Gallery~\citep{willis2021fusion} pairs a dataset of human design sequences with an environment in which a learned agent
rebuilds a target B-rep by face extrusions under random, beam or best-first search, and zone
graphs~\citep{xu2021zonegraphs} partition the target B-rep into volumetric zones and guide a search over extrusions and
Booleans with a graph neural network. On meshes, CADReasoner~\citep{kabisov2026cadreasoner} repeatedly edits a
complete program from its geometric discrepancy with the target, while CADENA~\citep{kabisov2026cadena} and
SOV-CAD~\citep{feng2026sovcad} predict one operation at a time from views of the target and of the current
reconstruction. CADENA trains its policy with supervised and reinforcement learning, samples several candidate
operations per step, and defines the CADENA-Bench benchmark of mechanical parts and the GMS metric of surface
agreement. Two methods add explicit search. CADFit~\citep{nehme2026cadfit} combines a learned sketch prior with
geometric fitting of extrusions to cross-sections of the target, assembles the program greedily, prunes it backward
and reconstructs the residual. StepCAD~\citep{nehme2026stepcad} trains a state-conditioned LLM policy on 1.5M
synthetic programs whose vocabulary includes lofts, sweeps, fillets and chamfers, and refines its output by an
IoU-guided beam search over four local edits: continuous refinement, replacing a sketch by a section of the mesh,
cutting away overshoot, and skipping an operation. In StepCAD's ablation, the search raises IoU from 0.682 to 0.816
over the policy alone.

Agents hand these decisions to a language model. CAD-Assistant~\citep{mallis2025cadassistant} plans calls to the
FreeCAD API with a vision-language model and CAD-specific tools, and IterCAD~\citep{hu2026itercad} generates and edits
CadQuery code over several turns from drawings and instructions. CADFather~\citep{savrasov2026cadfather} is the agent
closest to our setting. For a target mesh, a vision-language assistant keeps a pool of candidate programs and decides
which to extend and with which tool: samples from CADENA's policy, geometric proposals in the manner of CADFit, or a
parameter optimizer. It runs on GPUs within a limit of 1000~s per part and returns the best candidate by IoU and GMS.
Its ablations credit geometry: removing the geometric proposals lowers the mean score the most, by 0.012--0.110,
removing the optimizer costs 0.004--0.006, and the learned generator alone is the weakest configuration. On
BenchCAD~\citep{zhang2026benchcad}, frontier LLM agents with a Python sandbox self-report a voxel IoU of about 0.96 at
\$1.75--5.94 per part~\citep{benchcad2026leaderboard}. Like these methods, and unlike single-pass models, our method
evaluates candidates against the input mesh during inference, but it has no learned component: detection proposes and
evolution searches.

\paragraph{Parameter optimization.}
With the discrete structure of a program fixed, its continuous parameters can be optimized directly.
\citet{cascaval2022differentiable} make a CAD language differentiable to propagate direct geometric edits back to
program parameters, and DiffCSG~\citep{yuan2024diffcsg} differentiates the rasterization of CSG trees, which supports direct and image-based editing of primitive parameters. Among reconstruction methods, CADFather refines parameters with finite-difference gradients of a signed-distance mismatch, and StepCAD uses Bayesian optimization as one of its local edits. Our method uses gradient descent in the same role but inside the evolutionary loop: the genetic operators change the structure, and gradient descent tunes the continuous parameters of the better candidates in every
generation.

\section{Differentiable Geometry Kernel}
\label{sec:kernel}

The proposed method relies on its own geometric modeling kernel. 
It is not as general or as precise as industrial kernels such as Open CASCADE or Parasolid, which represent solids by their boundaries, 
but it implements exactly the operations that the program synthesis requires. 
It follows the constructive solid geometry (CSG) approach and represents a CAD model as a tree whose leaves are primitives and whose inner nodes are operations on them. 
The main operation is the evaluation, at a given point, of a scalar field whose negative set is the solid of the tree. 
For most nodes this field is the signed distance function (SDF) of the node's solid. 
For some nodes, such as the sweep, the exact SDF, or even a truncated SDF (TSDF), turned out to be too expensive to compute, so the kernel 
evaluates an approximation with the same zero set and the same sign. We call it a pseudo-SDF to avoid confusion. 
For some operations this function has larger Lipchitz bound (it is 1 for proper SDF) and can overestimate the distance away from surface. 
Importantly, this function is differentiable almost everywhere with respect to every continuous parameter of the tree. 
Most of our CSG trees can be triangulated exactly or with given precision via Manifold~\cite{manifold} library, while for general case we use 
manifold dual contouring~\citep{schaefer2007manifold} on an adaptive grid for approximate triangulation. 
Finally, this CSG tree representation could be converted into CadQuery program. 

Every node has a dimension: it describes a curve, a sketch (a planar region) or a solid, and it accepts children of the dimensions it expects. A node owns a short vector of parameters, and the parameters of a tree are those of its nodes in depth-first order. During optimization gradient descent changes this flat array of parameters while the structure of the tree is changed only by the structural operators of the search, e.g. mutations. \Cref{tab:nodes} lists the most common node types that the search uses.

\begin{table}[th]
	\caption{Node types of the kernel used by the search, with each node's own parameters.}
	\label{tab:nodes}
	\centering\small
	\begin{tabularx}{\linewidth}{@{}lllX@{}}
		\toprule
		dimension & node & parameters & meaning \\
		\midrule
		curve  & Line                       & --                                   & straight span between two consecutive points of a profile \\
			   & Spline                     & --                                   & Catmull--Rom spline between two consecutive points of a profile \\
		       & Arc, ArcInv                & $r_s > 0$                            & circular span with radius $\tfrac12\lVert q_{i+1} - q_i\rVert + r_s$, bulging to the right or to the left of its chord \\
		\hline
		sketch & Circle                     & center $c \in \mathbb{R}^2$, radius $r$ & disc \\
		       & Profile                    & points $q_1, \dots, q_n \in \mathbb{R}^2$ & closed loop with one curve child per span $q_i \to q_{i+1}$ \\
		       & Union, Subtract            & --                                   & 2D Booleans \\
		\hline
		solid  & Extrude                    & half-height $h$                      & sketch in the local $xz$ plane, extruded to $y \in [-h, h]$ \\
		       & Revolve                    & --                                   & sketch in the half-plane $(u, v)$, $u \ge 0$, turned a full turn around the local $y$ axis \\
		       & Sweep, SweepClosed         & knots $k_1, \dots, k_m \in \mathbb{R}^3$, span types & sketch swept along an open or a closed path \\
		       & Loft                       & half-height $h$                      & 2 sketches in the local $xz$ planes, mixed along $y \in [-h, h]$ \\
		       & Union, Subtract, Intersect & --                                   & 3D Booleans \\
		       & Translate3D, Rotate3D      & $t \in \mathbb{R}^3$, Euler angles $\alpha \in \mathbb{R}^3$ & rigid placement \\
		\bottomrule
	\end{tabularx}
\end{table}

\begin{figure}[th]
	\centering
	\includegraphics[width=0.75\linewidth]{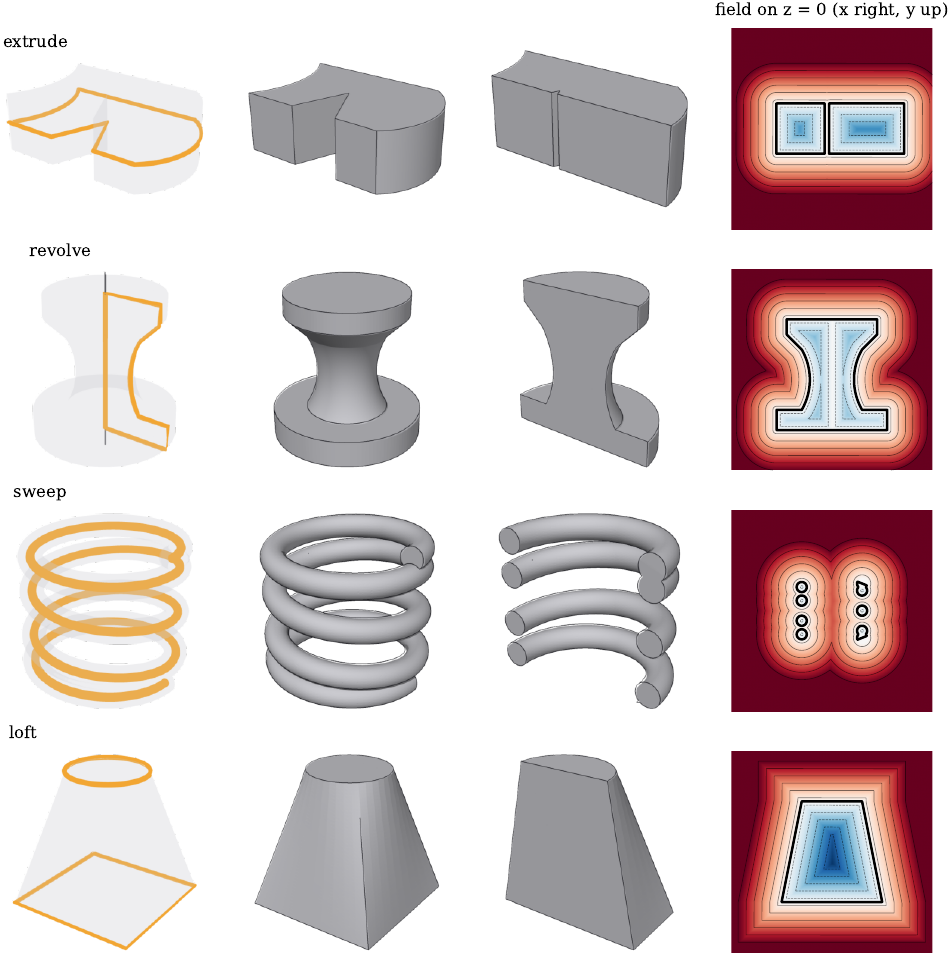}
	\caption{Example of main solid-creating operation: sketch, full solid, cut and pseudo-SDF visualization on cut plane.}
	\label{fig:kernel_features}
\end{figure}

\paragraph{Evaluation.}\label{sec:kernel_eval}
A tree is compiled once into a linear stream of operations and a flat array of parameters and its SDF is evaluated by a stack machine, as a few other CSG program executors do~\citep{sharma2018csgnet, grasberger2016efficient}. Operations are node-level and are of two types. A position operation transforms the current query point and pushes the result onto a stack: a 3D or 2D transform, or the projection of an extrude, a revolve or a sweep into the plane of its sketch. A distance operation
takes what the node's children left on a value stack and pushes their combination: the field of a leaf, a Boolean, or the join of an extrude's sketch field with its caps. Evaluation is a loop over the stream with two small stacks, with no virtual calls, no pointers and no memory allocation, and the parameters can change between evaluations without recompiling the stream, which is what gradient descent does between its steps. The stream is the same for every query point, so a batched evaluator applies each operation to 8 (AVX) or 16 (AVX-512) points at once, and expresses data-dependent choices, such as the nearest curve of a profile, by masks. 

\paragraph{Derivatives.}\label{sec:kernel_diff}
Gradient descent (\cref{sec:gd}) needs $\partial f / \partial \theta$ for every parameter $\theta$ of the tree at a batch of points. A second stack machine computes it together with the field, in one pass over the same stream. Each entry of its value stack holds the field of a subtree and the field's gradient with respect to the query point in the subtree's frame, and the parameter gradients are written into one flat array indexed like the parameters. A leaf writes the gradient of its field with respect to its own parameters. A transform leaves the parameter gradients of its subtree as they are, since its child was evaluated at the transformed point, computes its own by the chain rule through that point, and maps the point gradient back into its parent's frame. A Boolean keeps the gradients of the child whose value it took and zeroes those of the other one, and a scale multiplies the gradients of its subtree by its factor. The resulting scheme is a mixed-mode algorithmic differentiation, forward derivatives calculation over the distance tree, reverse over the transform chain. It also has AVX and AVX-512 batched evaluator.

\section{Optimization Pipeline}
\label{sec:method}

\begin{figure*}[th]
	\centering
	\includegraphics[width=\linewidth]{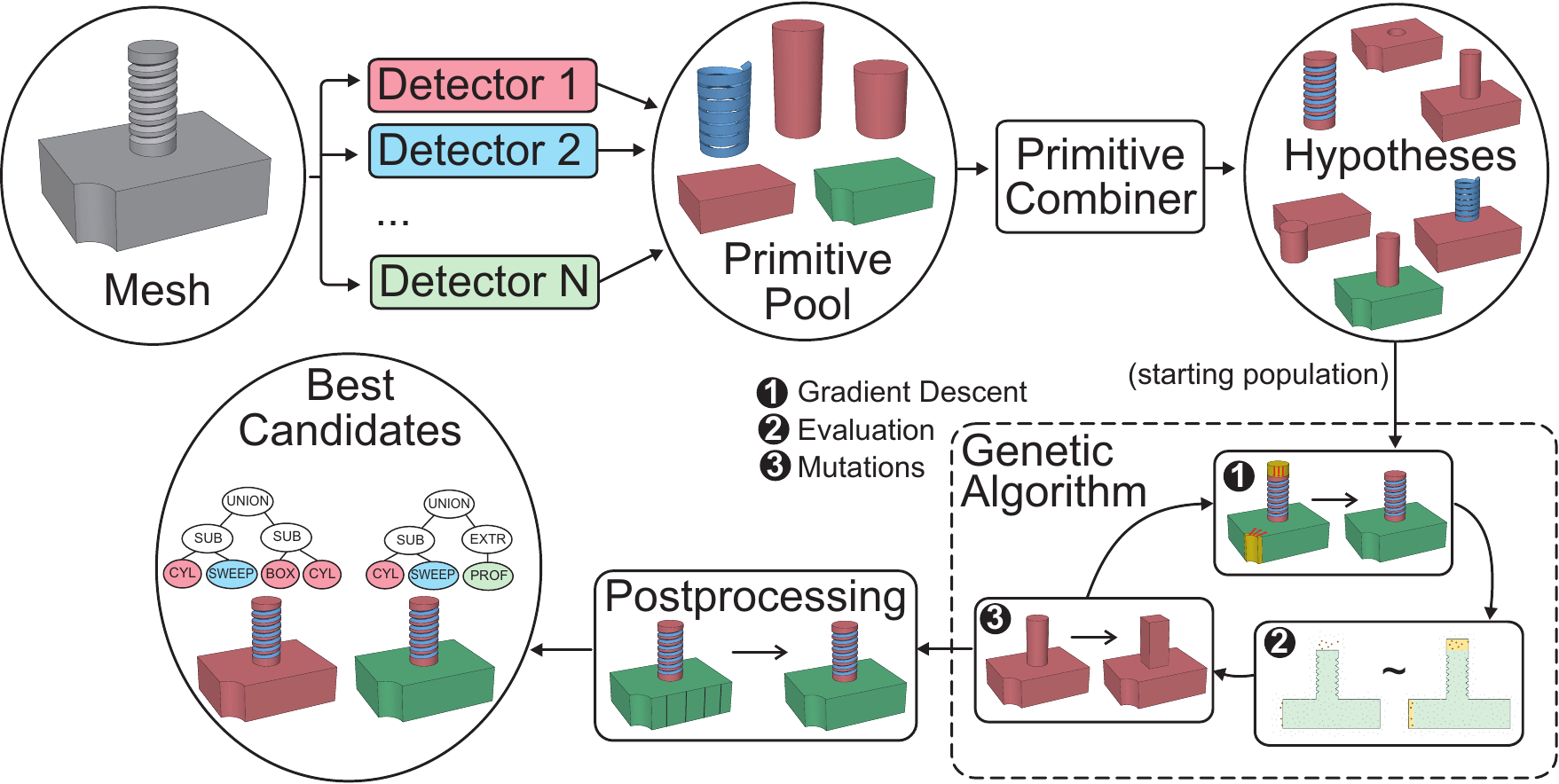}
	\caption{Overview of the method. The target mesh is analyzed by a few different \textit{primitive detectors}, each finding its own set of primitives and contributing them to the pool. \textit{Primitive combiner} then assembles hypotheses, CSG trees consisting of primitives from the pool. These hypotheses are used as a staring population for \textit{genetic algorithm}, which evolves them for many generations. On each generation population is first improved via gradient descent, then evaluated against target mesh and new generation is created via crossover and mutation of the best ones. After the target score has been reached or evaluation budget exhausted, a few best scored CSG candidates are post-processed, which includes simplification of tree structure and parameters fine-tuning. After post-processing these candidates are presented to the uses as CAD programs (e.g. via export to CadQuery).}
	\label{fig:pipeline}
\end{figure*}

We propose a CAD program synthesis pipeline (\cref{fig:pipeline}) that detects primitives in the input mesh,
combines them into hypotheses (complete candidate programs) and uses these as the initial population of a genetic
algorithm. The genetic algorithm searches the space of programs with mutation and crossover operators that change
their structure, and with a local search by gradient descent that tunes their continuous parameters. When the target
quality is reached or the computational budget is spent, the best candidates are refined at a post-processing stage,
which simplifies the programs without sacrificing quality. Some stages, such as the local search, are mostly fixed;
others, primitive detection above all, are open and can take in components of other mesh-to-CAD methods, as
\cref{sec:ablation} shows for CADFit~\cite{nehme2026cadfit} and CADFather~\cite{savrasov2026cadfather}. After the problem setting (\cref{sec:setting}), each main block of the algorithm is described in its own subsection.

\paragraph{Problem Setting.}\label{sec:setting}
The input for the method is a triangle mesh. It is repaired into the boundary of the solid $T$ that it encloses
(\cref{sec:preprocessing}), then scaled uniformly and centered to fit in the cube $[-1, 1]^3$. The output is a program $P$ whose solid $S(P)$ matches $T$, and the objective is the intersection over union (IoU). The target is fixed while many candidates are compared with it, so everything the search
needs to know about $T$ is precomputed before the search for performance reasons. The fitness of a program is an estimate of its IoU with the target, computed by an hierarchical estimator with adaptive precision. It provides coarse estimation for bad candidates and precise one to those very similar to target, allowing efficient genetic optimization, see~\cref{sec:fitness} for details.

\subsection{Primitive Detection}
\label{sec:detection}

The goal of a primitive detector is to find, in the input mesh, solids that are likely to be features of the CAD program. 
In general, a primitive is any valid CSG tree; what exactly a primitive can be depends on the detector. Each primitive carries a CSG sign (solid or cut), which says whether it is meant to be added to the part or subtracted from it; a confidence; and a set of \textit{conflict keys}. Conflict keys are a general way for a detector to express that some primitives are rival explanations of the same shape. For example, an $N$-sided prism in the mesh can be interpreted literally, as an extrude of an $N$-sided polygon, or as a cylinder, so a detector adds both, but with the same conflict key. Two primitives sharing a conflict key never appear in one hypothesis. This feature allows detectors to emit alternatives rather than decide between them, and leave the decision to the combiner and the search.

Primitive detection is the open and extensible part of the pipeline: several detectors run on the same mesh, and
adding one aimed at a specific class of shapes needs no change elsewhere. We created two detectors, \textbf{RANSAC-extrude} and \textbf{RANSAC-direct}, and also demonstrate the integration of CADFit~\citep{nehme2026cadfit} and CADFather~\citep{savrasov2026cadfather} as primitive detectors (\cref{sec:ablation}). 
Both of our detectors use one tool, which extracts shapes from the mesh in the manner of efficient RANSAC~\citep{schnabel2007ransac} but on triangles instead of points~(\cref{sup:detectors}). A candidate shape is fitted to a seed triangle, or to a small patch around it; its support grows over the connected triangles that lie
within a band of half-width $\epsilon$ around the shape and whose normals agree with it; the best of many candidates is
extracted, its triangles are retired, and the extraction continues until most of the triangles are gone or draw limit reached.

\textbf{RANSAC-extrude} detects only planes and cylinders, and assembles extrudes from them (\cref{sup:extrude_combiner}). A full cylinder is a primitive by itself, two parallel planes facing away from each other are the caps of an extrude whose profile is the outline of their triangles. A plane with no opposite cap is the floor of a cut of unknown depth, extruded outwards. \textbf{RANSAC-direct} detects extrudes, revolves and sweeps directly as shapes of the engine, alongside planes. Each shape becomes a primitive whose polarity comes from a vote of the normals of its walls; when the vote is unclear, the detector emits both a solid and a cut with a shared conflict key. Together, primitive detectors are a crucial part of the pipeline: they extract much of the information from the input geometry, and together with the primitive combiner they let the evolutionary loop start from a better position. For relatively simple parts, the detectors find all features outright.

\subsection{Primitive Combiner}
\label{sec:combiner}
The primitives of all detectors form a pool. For every primitive the pool stores, as a bitset, which of the target's fixed volume samples lie inside it, and from these bitsets the pairwise overlaps of the primitives. From then on, every comparison between primitives and every IoU of a combination of them is a count of bits on one fixed set of points: it is free of sampling noise between candidates and fast enough for thousands of combinations.

The combiner turns the pool into hypotheses: subsets of the pool, each converted into a program that unites the
subset's solids in the order of detection and then subtracts its cuts. It first builds a conflict graph. Two
primitives from one detector conflict if they share a conflict key, and two primitives of the same polarity conflict
if one covers more than 90\% of the other's volume, since such a pair is usually two versions of one feature. A
hypothesis should contain no conflicting pair and should not leave out a primitive that could join it without a
conflict, so the combiner enumerates the maximal conflict-free sets, that is, the maximal independent sets of the
conflict graph. They are found by the Bron--Kerbosch algorithm~\citep{bron1973cliques} on the complement of each
connected component of the graph and combined across the components. When a component has too much of such
sets, the enumeration is replaced by a fixed number random maximal sets, each built greedily in a random order of the
primitives. When there are fewer sets than hypotheses requested, the combiner adds subsets of them with one, two or more primitives left out, largest first. Every solid primitive is also a hypothesis of its own, because when everything else fails one large primitive is often a fair start.Each set is then pruned greedily for IoU on the pool's samples: the primitive whose removal increases the IoU the most is removed, until no removal increases it, and the last solid is never removed. Pruning drops primitives that are useless or harmful in one particular combination while keeping them in others.

\subsection{Genetic Algorithm}
\label{sec:ga}

\begin{algorithm}[ht]
	\caption{Evolutionary search on one target}
	\label{alg:ga}
	\begin{algorithmic}[1]
		\Require target data, pool, hypotheses $H$
		\State $\mathit{pop} \gets$ the best $N$ of $H$ and random programs, by precise fitness
		\State $\mathit{best} \gets$ the $N_{best}$ best programs of $\mathit{pop}$
		\While{evaluations $<$ budget}
		\State find look-alikes by occupancy signature
		\ForAll{$P \in \mathit{pop}$ that are not look-alikes} \Comment{in parallel}
		\State $F(P) \gets \widehat{\mathrm{IoU}}_{\mathrm{coarse}}(P)$
		\EndFor
		\ForAll{those $P$ with $F(P)$ above the median} \Comment{in parallel}
		\State $P \gets \mathrm{GradientDescent}(P)$;\quad $F(P) \gets \widehat{\mathrm{IoU}}_{\mathrm{precise}}(P)$
		\EndFor
		\State sort $\mathit{pop}$ by $F$; update $\mathit{best}$ from the precisely evaluated programs
		\If{$F(\mathit{best}_1) \ge F^\ast$ and an independent precise estimate agrees}
		\State \textbf{break}
		\EndIf
		\If{the best fitness stagnates}
		\State rebuild $\mathit{pop}$ from new hypotheses and random programs, with $\mathit{best}$ copied in
		\Else
		\State $\mathit{next} \gets$ the best $E$ programs of $\mathit{pop}$
		\ForAll{remaining places} \Comment{in parallel}
		\State $C \gets$ a random program, or $\mathrm{Mutate}(\text{crossover or copy of tournament winners})$
		\State draw again while $C$ is not valid
		\EndFor
		\State $\mathit{pop} \gets \mathit{next}$
		\EndIf
		\EndWhile
		\State \Return $\mathrm{PostProcess}(\mathit{best})$
	\end{algorithmic}
\end{algorithm}

The search keeps a population of programs and runs generations until it spends a given number of fitness evaluations
or the best program reaches the target fitness. \Cref{alg:ga} summarizes one run.

\paragraph{Initial population.}
The valid hypotheses of the combiner are completed by random programs if there are fewer than $N$ of them. A random
program is a chain of one to three random extrudes or revolves, each placed inside the bounding box of the previous
one and cut from the part with probability 0.3. All of them are evaluated precisely, and the best $N$ form the first
generation.

\paragraph{Look-alikes.}
Many programs differ only in ways that do not change their shape near the target surface, typically copies of an elite 
with neutral shape-preserving mutations. Before evaluation of the population, we sample occupancy (SDF sign) for each
program at a set of $K$ points ($K = 4096$  in experiments) from the near-surface samples and use it as a $K$-bit 
signature. Programs with equal signatures are look-alikes: only the first of them is evaluated, and the others take its fitness minus $10^{-5}$, so that an evaluated program always ranks above its copies.

\paragraph{Evaluation.}
Every other program gets a coarse fitness. The programs above the median of the coarse fitness (in the first generation, all of them) are tuned by gradient descent (\cref{sec:gd}) and then evaluated precisely; the precise value replaces the coarse
one. The population is sorted by fitness. A list of the few best programs of the run is updated from the precisely evaluated programs only, since coarse estimates are too noisy to compete with precise ones. The best $E$ programs are copied into the next generation unchanged. They are screened and refined again there, so the best programs are tuned over many generations, a few steps at a time. When the best program reaches $F^\ast$ for the first time, it is evaluated once more, precisely and on different sample points, and the run stops if that estimate reaches $F^\ast$ too. This keeps a lucky estimate from ending the run early. If the best fitness has gained almost nothing over several generations and the budget has room for more, the population is rebuilt as at the start, with the $N_{best}$ best programs copied in.

\paragraph{Offspring.}
The remaining $N - E$ places are filled with new programs. With small probability a new program is random. Otherwise, with probability 0.5, it is the crossover of two parents, and otherwise a copy of one; either way one mutation is then applied to it. Parents are chosen by tournaments. 

\paragraph{Crossover.}
Crossover is one-point on the spines: the child takes the first $k$ steps of one parent and the steps from $k$ on of
the other, with $k$ uniform up to the length of the longer spine and the roles of the parents drawn at random. Every
gene is placed in the frame of the scene, not relative to the steps before it, so a step keeps its meaning in the
child: a hole taken from one parent still cuts where it cut before.

\begin{figure}[th]
	\centering
	\includegraphics[width=0.95\textwidth]{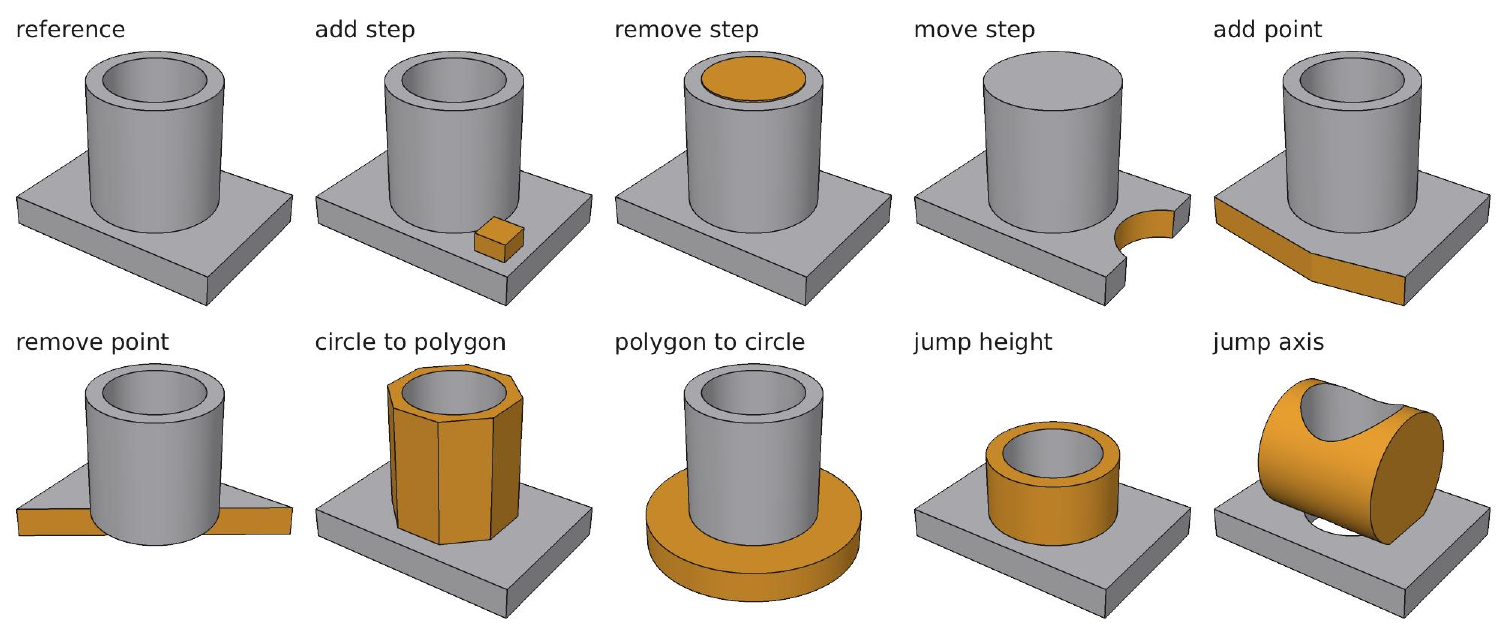}
	\caption{Examples of different mutations on programs.}
	\label{fig:operators}
\end{figure}

\paragraph{Mutations.}
Mutations are drawn by weight from a table (\cref{tab:mutations}, \cref{sup:mutations}). Structural mutations add a
step anywhere in the spine, remove one, move one to another position, flip a union into a cut and back, or swap a
gene that came from the pool for a rival primitive that shares a conflict key with it. An added step is a pool
primitive or a random gene placed inside what the spine builds. Frame mutations move a gene, turn it (among right angles if it is locked), or lock and unlock its frame. Sketch mutations add or remove a profile point, add or remove a hole, turn a circle into a regular polygon of the same area and back. Feature mutations change the height of an extrude or the axis of a revolve, or convert an extrude of concentric circles into the equivalent revolve of rectangles and back. Adding and removing profile points have by far the largest weights. A new point splits a line at its midpoint, so it does not change the shape until gradient descent moves it; together, these two mutations and gradient descent are how profiles adapt their shape.
After a mutation the program is repaired, with angles wrapped, profiles turned counter-clockwise and parameters
clamped to their limits, and then checked for validity. \Cref{fig:operators} shows examples of different mutations applied to one program.

\paragraph{Determinism.}
Every random decision draws from a seed derived from the run's seed, the kind of decision, the generation and the
index of the program, and never from the state of a thread. A run therefore gives the same result on any number of
threads, so a single part can be optimized on all cores of a machine (evaluation, refinement and offspring creation
are parallel loops over the population) without losing reproducibility.

\subsection{Gradient Descent}
\label{sec:gd}

Genetic operators are good at changing structure but poor at tuning continuous parameters: a random jump of a
coordinate almost never improves a program that is already close to the target. Every program that passes the
screening is therefore tuned by gradient descent before its precise evaluation. On a fixed set of samples IoU is
piecewise constant in the parameters. We fit the program's field to the target's signed distance near the surface instead. With the near-surface samples $p_j$ and their signed distances $d_j$ to the target (\cref{sec:setting}), the loss is
\begin{equation}
	L(\theta) = \frac{1}{|\mathcal{B}|} \sum_{j \in \mathcal{B}} \bigl(f_P(p_j; \theta) - d_j\bigr)^2
	\label{eq:gd_loss}
\end{equation}
over a random batch $\mathcal{B}$ of samples in each step. Its minimum puts the zero set of $f_P$ on the target's
surface, and since the samples reach $d_{near}$ into both sides, a surface that is off by less than that is pulled towards
the right place. We run a few dozen steps of Adam~\citep{kingma2015adam} with the analytic gradients of
\cref{sec:kernel_diff}, and after every step clamp the parameters to their box constraints. The learning rate depends on the program's fitness $F$: it is $\eta(F) = 2 \cdot 10^{-4}\, m(F)$, where $m$ falls from 2 for the poorest programs to 1 at $F = 0.75$ and to 0.1 at $F = 1$ (\cref{sup:params}), so that good programs are polished rather than significantly changed.

Gradient descent changes neither the structure of a program nor the limits of its parameters, but it can make a loop
intersect itself or fold the path of a sweep onto itself. After the descent, every loop that intersects itself, and
every sweep gene whose path became invalid, gets back its parameters from before the descent; all other parameters
keep their new values.

\begin{figure}[th]
	\centering
	\includegraphics[width=\linewidth]{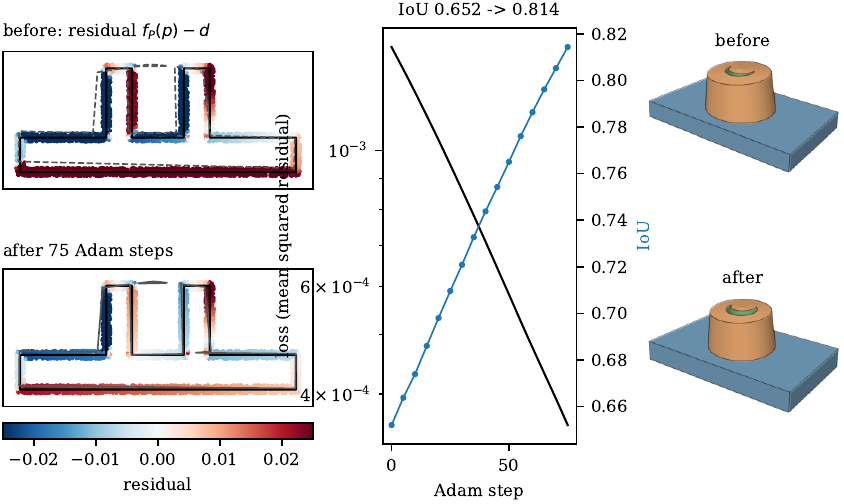}
	\caption{Local search fits the program's field to the target's signed distance near its surface. A program with
		its parameters perturbed, before and after one call of 75 Adam steps. Left: a slice with the target's section
		(solid), the program's zero set (dashed) and the near-surface samples coloured by the residual
		$f_P(p_j) - d_j$. Middle: loss and IoU against the step. Right: renders.}
	\label{fig:gd}
\end{figure}

\subsection{Post-processing}
\label{sec:postprocess}

The search optimizes IoU only, so its programs can carry genes that contribute almost nothing, and profiles with
more points than their shapes need. The post-process simplifies $N_{best}$ programs of the run without giving up
accuracy. Every edit is judged on one fixed set of points, uniform in the target's bounding box enlarged slightly on every side, on which the occupancy of every gene is computed once. Boolean combination of bitsets, and its IoU with the target is estimated just by counting bits. A removed gene stands for the empty set where it was: $A \cup \emptyset = A \setminus \emptyset = A$, while $\emptyset \setminus A$ and $A \cap \emptyset$ are empty and take their other operand with them.

\paragraph{Dead genes.}
The genes are tried for removal from the smallest to the largest, by the number of points inside them. A gene is
removed if the program without it, and without the genes removed before it, loses at most $\epsilon_{F}$ IoU against the full
program, so that the losses cannot add up beyond that.

\paragraph{Profiles.}
The profiles of the remaining genes are then simplified. A profile made only of lines that fits a circle to within
tolerance $\epsilon_{A}$ of area, becomes a circle by the algebraic fit \citet{kaasa1976circle} measured at its points and at the  midpoints of its segments. Any other profile is simplified by an arc-aware algorithm that replaces runs of
consecutive curves by single lines or arcs. An arc between two pinned points has one free parameter, so fitting one to a run is a one-dimensional problem with a closed form, and the fitted arc stays connected to its neighbours. All runs are fitted once, and dynamic programming over the closed loop, started at every vertex in turn, chooses the subset of runs with the fewest curves whose deviation from the original outline stays below $\epsilon_{A}$ of the profile's size. A simplified gene is kept if the program loses at most $\epsilon_{F}$ IoU; otherwise it is tried again with the tolerances halved, up to three times.

\paragraph{Acceptance.}
The simplified program replaces the original if it can be meshed into a valid solid and loses less than $\epsilon_{F}$ of
precise IoU. Of the two programs, the one with the higher precise IoU is the result.

\begin{figure}[th]
	\centering
	\includegraphics[width=\linewidth]{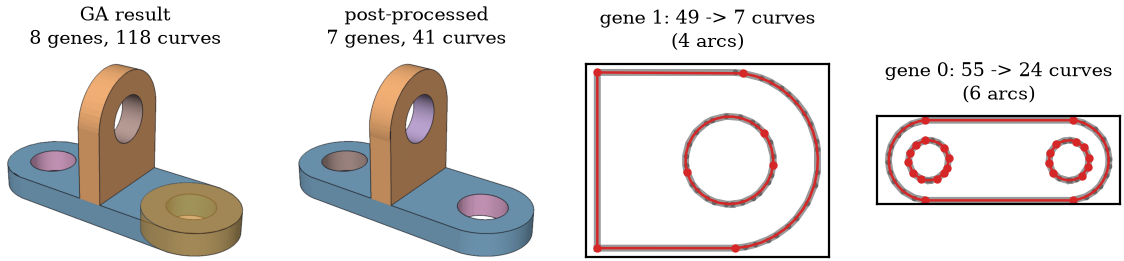}
	\caption{Post-processing removes genes that carry no volume and replaces polylines by lines, arcs and circles. Left: the program found by the search and the post-processed one, coloured by gene. Right: two profiles before (grey polyline) and after (red: lines and arcs, end points as dots).}
	\label{fig:postprocess}
\end{figure}

\section{Implementation Details.}
The method is implemented in C++ and runs on CPUs only. A run stops when the fitness of its best program reaches $F^\ast = 0.995$ or after $50'000$ fitness evaluations; there is no time limit. Unless stated otherwise, every part runs on a single thread, with one part per hardware thread of a 16-core desktop CPU (Ryzen 9 7950X), so our times are thread-seconds on a fully loaded machine. All settings are listed in \cref{sup:params}. All experiments across all datasets were run with this exact settings.

\begin{figure}[th]
	\centering
	\includegraphics[width=\linewidth]{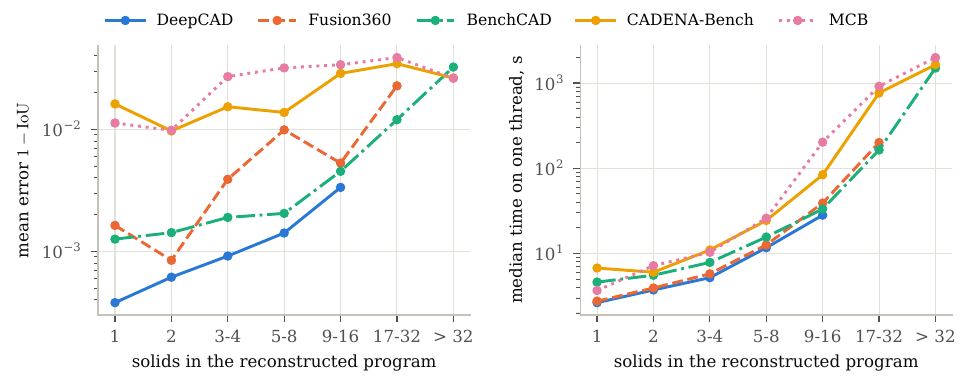}
	\caption{Mean error (left) and median time (right) against the number of solids in the reconstructed program.
		Groups with fewer than 20 parts are omitted.}
	\label{fig:complexity}
\end{figure}

\Cref{fig:distributions} (right) shows the distribution of the time per part on one CPU thread, and
\cref{fig:complexity} relates it to the size of the reconstructed program. A DeepCAD part takes 3.3~s in the median
and a Fusion360 part 4.6~s. On mechanical parts the median is a few minutes (205~s on MCB and 259~s on CADENA-Bench), and the
distribution has a long tail: 31\% of the MCB parts and 32\% of the CADENA-Bench parts take more than 1000~s on one thread, which is
CADFather's limit per part. On BenchCAD, the median is 11~s, and five families (coil springs, twist drills, pipe elbows, helical and bevel gears) take 71\% of the total time.
The search parallelizes over the population (\cref{sec:ga}), and a run gives the same program on any number of threads. 
On all 32 threads of a desktop CPU, it leads to 10-20 times speed-up compared to single-threaded run.

\section{Results}
\label{sec:experiments}

We have validated our method on variety datasets to compare it against a wide range of existing mesh-to-CAD methods since there is no single dataset that brings together all the work in this field. So, we have used the full test sets of DeepCAD~\citep{wu2021deepcad} (8046 parts), Fusion360~\citep{willis2021fusion} (1725) and MCB~\citep{kim2020mcb} (5000); CADENA-Bench~\citep{kabisov2026cadena}, (3396 mechanical parts in six families); and all 17900 parts of BenchCAD~\citep{zhang2026benchcad}, from 106 families in three difficulty tiers. We also used 50-model dataset of InverseCSG~\citep{du2018inversecsg} specifically to compare with this algorithm. \Cref{tab:main} compares our method with existing work.

\paragraph{Target preprocessing.}\label{sec:preprocessing}
The meshes in used datasets are exports of CAD models, and many of them are not boundaries of solids: some are open, some consist of overlapping or touching shells, and some contain internal walls. Volumetric IoU is not defined for such a mesh, and neither is our fitness. Before reconstruction, every input mesh is therefore repaired into the boundary of the solid that it encloses. Holes are capped, all triangles are cut along their mutual intersections by an exact mesh arrangement, and the winding number of the closed surface decides which pieces separate the inside from the outside; internal faces and sealed cavities are removed, and cracks are sealed. The outcome has one of five labels: \emph{not modified} (the mesh already is such a boundary), \emph{cleaned up} (faces were removed, none was changed), \emph{repaired} (faces were flipped or cut, the solid is
what the input encloses), \emph{completed} (caps remain in the result, so part of the surface is inferred) and
\emph{rejected} (the mesh encloses no volume, or the repair fails). \Cref{tab:targets} gives their counts. The
repaired mesh is the target of the search and of every metric; a rejected target counts as a failure of the method.

\begin{table}[th]
	\caption{Targets after preprocessing. The last column counts the targets that are one closed solid with positive
		volume, the targets on which other methods report IoU.}
	\label{tab:targets}
	\centering\small
	\begin{tabular}{@{}lrrrrrrr@{}}
		\toprule
		benchmark & parts & not modified & cleaned up & repaired & completed & rejected & one closed solid \\
		\midrule
		DeepCAD      & 8046  & 8004  & 1   & 15   & 24  & 2  & 7651 \\
		Fusion360    & 1725  & 1583  & 7   & 131  & 4   & 0  & 1604 \\
		MCB          & 5000  & 2693  & 13  & 1865 & 405 & 24 & 3943 \\
		CADENA-Bench & 3396  & 2067  & 164 & 1022 & 126 & 17 & 2864 \\
		BenchCAD     & 17900 & 17157 & 50  & 505  & 187 & 1  & 16610 \\
		\bottomrule
	\end{tabular}
\end{table}

\paragraph{Metrics and protocol.}
We follow the protocol of CADFather~\citep{savrasov2026cadfather}, which extends that of
CADENA~\citep{kabisov2026cadena}. IoU is the exact volumetric IoU of the target and the mesh of the result, computed with mesh Booleans~\citep{manifold}; it is independent of the estimator that the search optimizes (\cref{sec:fitness}). GMS measures the agreement of the two surfaces and their normals, normalized to $[0, 1]$, and the Chamfer distance (CD) is computed on two different points count: 8192 and on 30000 points per surface. IoU and GMS are reported as means and CD as a median times $10^3$. For our method, all means and medians are taken over all targets that were not rejected; the few rejected ones count in invalidity ratio (IR). The result is built in the frame of the target, so nothing is aligned or rescaled before measuring. All metrics of our method are computed on the meshes created by our the kernel.

\begin{table}[th]
	\caption{Comparison with state-of-the art methods on multiple datasets. Rows of other methods are published results~\citep{kabisov2026cadena,savrasov2026cadfather}. 
	The invalidity ratio (IR) is the share of parts without a valid output for each method. 
	IoU and GMS is a means, CD is a median times $10^3$ (for 8192 and on 30000 points per surface according to \citep{savrasov2026cadfather} test protocol), 
	and time the mean time per part in seconds: on one GPU for CAD-Recode and cadrille~\citep{kolodiazhnyi2025cadrille}, as reported by the authors for CADFit and CADFather. 
	$^\ast$\, mean times of our method are in single-threaded mode, multi-threaded reconstruction is 10-20 times faster on 32-thread CPU. 
	$^\dagger$\, mean time on a subset of complex parts~\citep{nehme2026cadfit}.}
	\label{tab:main}
	\centering\small
	\begin{tabular}{@{}llrrrrrr@{}}
		\toprule
		dataset & method & IR (\%) & IoU & GMS & CD$_{8k}$ & CD$_{30k}$ & time, s \\
		\midrule
		DeepCAD      & CAD-Recode          & 7.90\%  & 0.914 & 0.929 & 0.17  & 0.055 & --  \\
		             & cadrille            & 0.30\%  & 0.897 & 0.948 & 0.17  & 0.059 & --  \\
		             & CADENA              & 0.07\%  & 0.966 & 0.974 & 0.15  & 0.042 & -- \\
		             & CADFit              & 27.50\% & 0.914 & 0.898 & 0.18  & 0.052 & 242$^\dagger$ \\
		             & CADFather           & 0.00\%  & 0.987 & 0.983 & \textbf{0.138} & \textbf{0.039} & 82.6 \\
		             & Ours                & 0.02\%  & \textbf{0.999} & \textbf{0.987} & 0.142 & 0.040 & \textbf{10.9} \\
		\midrule
		Fusion360    & CAD-Recode          & 13.60\% & 0.872 & 0.858 & 0.17  & 0.060 & -- \\
		             & cadrille            & 0.80\%  & 0.848 & 0.868 & 0.17  & 0.066 & -- \\
		             & CADENA              & 0.23\%  & 0.952 & 0.942 & 0.12  & 0.036 & -- \\
		             & CADFit              & 25.60\% & 0.882 & 0.846 & 0.17  & 0.054 & 235 \\
		             & CADFather           & 0.00\%  & 0.976 & 0.960 & \textbf{0.115} & 0.034 & 119.9 \\
		             & Ours                & 0.00\%  & \textbf{0.996} & \textbf{0.973} & \textbf{0.115} & \textbf{0.032} & \textbf{24.6} \\
		\midrule
		MCB          & CAD-Recode          & 21.50\% & 0.711 & 0.547 & 0.66  & 0.524 & -- \\
		             & cadrille            & 1.50\%  & 0.662 & 0.550 & 0.85  & 0.720 & -- \\
		             & CADENA              & 0.06\%  & 0.895 & 0.751 & 0.22  & 0.089 & -- \\
		             & CADFit              & 29.70\% & 0.758 & 0.601 & 0.48  & 0.365 & -- \\
		             & CADFather           & 0.00\%  & 0.913 & 0.766 & 0.210 & 0.083 & 245.6 \\
		             & Ours                & 0.48\%  & \textbf{0.972} & \textbf{0.858} & \textbf{0.155} & \textbf{0.048} & 1019$^\ast$ \\
		\midrule
		CADENA-Bench & CAD-Recode          & 27.90\% & 0.686 & 0.489 & 1.061 & 0.905 & -- \\
		             & cadrille            & 0.80\%  & 0.688 & 0.498 & 1.267 & 1.125 & -- \\
		             & CADENA              & 1.60\%  & 0.876 & 0.670 & 0.331 & 0.176 & -- \\
		             & CADFit              & 18.80\% & 0.757 & 0.548 & 0.713 & 0.536 & -- \\
		             & CADFather           & 0.00\%  & 0.909 & 0.721 & 0.293 & 0.136 & 306.7 \\
		             & Ours                & 0.50\%  & \textbf{0.975} & \textbf{0.853} & \textbf{0.202} & \textbf{0.062} & 1117$^\ast$ \\
		\bottomrule
	\end{tabular}
\end{table}

\begin{figure}[th]
	\centering
	\includegraphics[width=\linewidth]{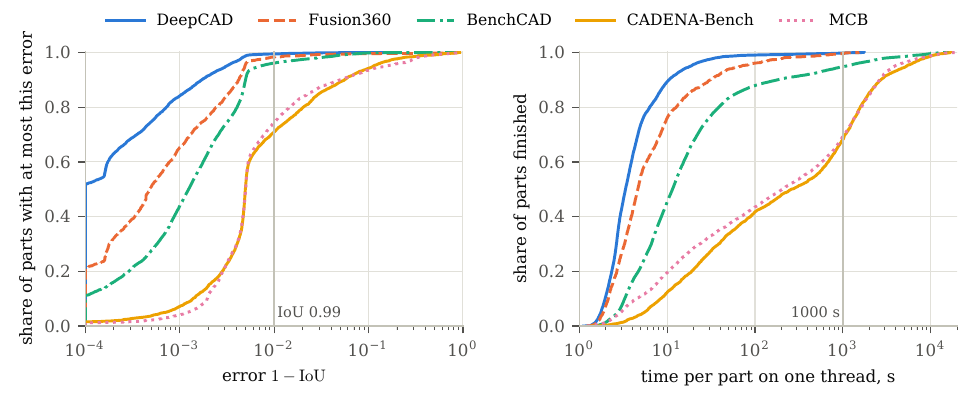}
	\caption{Distribution of the error (left) and of the time per part on one thread (right) over the parts of each
	benchmark. The step of the error curves at 0.005 is the stopping criterion $F^\ast = 0.995$.}
	\label{fig:distributions}
\end{figure}

\paragraph{DeepCAD and Fusion360.}
The mean IoU over all targets is 0.999 and 0.996, against 0.987 and 0.976 for CADFather, the best published results,
so the remaining error $1 - \mathrm{IoU}$ is a sixth of CADFather's or less. 99.6\% of the DeepCAD targets and 98.8\% of
the Fusion360 targets are reconstructed with an IoU of at least 0.99 (\cref{fig:distributions}, left). GMS is also higher
than CADFather's. The median CD matches CADFather's on both benchmarks. 
It has reached the noise of the protocol there, which samples the two surfaces independently: the 4165 DeepCAD results with an IoU above 0.9999 have a median CD of $0.150 \cdot 10^{-3}$ at 8192 points.

\paragraph{MCB and CADENA-Bench.}
MCB and CADENA-Bench have much more complex models, where the published methods lose most of their accuracy, and where the differences between methods are large enough to measure. 
On CADENA-Bench, the mean IoU over all targets is 0.975 against 0.909 for CADFather and 0.876 for CADENA, so the remaining error is less than a third of CADFather's.
GMS is 0.853 against 0.721, and the median CD at 30000 points is less than half. 
77\% of its single-solid targets (71\% overall) are reconstructed with an IoU of at least 0.99. 
On MCB, the mean IoU is 0.972 over all targets (likewise less than a third of CADFather's remaining error at 0.913) and 0.979 on those that are one closed solid.
GMS is 0.858 against 0.766, the median CD at 30000 points is 0.048 against 0.083, and 80\% of its single-solid targets (74\% overall) reach an IoU of at least 0.99. 
On both benchmarks, a large share of meshes required repair (\cref{tab:targets}), with 17 targets (0.50\%) rejected on CADENA-Bench and 24 (0.48\%) on MCB.
\paragraph{Time.}
The last column of \cref{tab:main} gives the mean time per part. On one CPU thread, our method is faster than the
other methods that optimize against the target on DeepCAD and Fusion360 (10.9~s and 24.6~s against 83~s and 120~s
for CADFather) and slower on mechanical parts (1019~s against 246~s on MCB and 1117~s against 307~s on CADENA-Bench), where a few complex parts
dominate the mean. This is due to design of experiment, where only one thread was allocated per part.

\paragraph{BenchCAD.}
BenchCAD was published after most of the methods above, and frontier language models with a Python sandbox are
evaluated on it. \Cref{tab:benchcad} compares our results with them and also CADENA~\citep{kabisov2026cadena} and CADFather~\citep{savrasov2026cadfather}. Our method again scores much higher: mean IoU is 0.997 against 0.948 for CADFather;
A part takes 11~s in the median and 245~s on average on one thread of a desktop CPU, which costs a fraction of a cent in electricity, against \$0.046 in model tokens for CADFather and \$1.75--5.94 in API cost for the agents. However, we should note a limitation of this comparison, as frontier models were not provided with the mesh and were solving potentially harder image-to-CAD task.

\begin{table}[th]
	\caption{Comparison on BenchCAD dataset. IoU is the exact IoU; voxel IoU (on a $64^3$ grid) and IR (the share of programs that fail to execute) are the metrics of the benchmark's own protocol. 
	Other rows are published results taken from~\citep{benchcad2026leaderboard,savrasov2026cadfather}. $^\ast$\,self-reported on a random 1000-part subset. 
	$^\dagger$\,IoU on all parts, voxel IoU and cost on a random 1000-part subset. 
	Cost is per part: API cost for the agents (provided as of the end of 2026), model tokens for CADFather, and for our method the electricity for a few core-minutes on a desktop CPU.}
	\label{tab:benchcad}
	\centering\small
	\begin{tabular}{@{}lllrrrr@{}}
		\toprule
		method & input & parts & IR (\%) & IoU & voxel IoU & cost per part \\
		\midrule
		Claude Opus 5.5 (max)$^\ast$ & images + tools & 1000  & --      & --    & 0.962  & \$5.94 \\
		GPT-6 Astra$^\ast$           & images + tools & 1000  & --      & --    & 0.959  & \$1.75 \\
		CADENA                       & mesh           & 17900 & 0.9     & --    & 0.910  & -- \\
		CADFather$^\dagger$          & mesh           & 17900 & 0.0     & 0.948 & 0.968  & \$0.046 \\
		Ours                         & mesh           & 17900 & $<$ 0.1 & \textbf{0.997} & \textbf{0.989} & $<$ \$0.01 \\
		\bottomrule
	\end{tabular}
\end{table}

\paragraph{Comparison with InverseCSG.}
InverseCSG~\citep{du2018inversecsg} is the closest method that is not learning-based, and it argued against
evolutionary search (\cref{sec:related}). We run our method on its 50 models with the configuration of all other
experiments, and compare the relative volumetric error per model with the published one (\cref{tab:icsg}). The two methods are equally accurate given the small size of the dataset: the mean error is 0.83\% for ours and 0.89\% for InverseCSG, the medians are 0.46\% and 0.40\%, and our error is the lower one on 28 of the 50 models. The main difference is in the time. InverseCSG reports 38.3 minutes per model on average and 330 minutes for the slowest; our method takes 21.5~s on average and 202~s for the slowest, on all threads of a desktop CPU. Even counted as thread-time on all 32 threads, this is 11.5 minutes on average. The models are as complex as those of CADENA-Bench (20 solids per program on average), so the genetic search is not the bottleneck that InverseCSG found it to be.

\begin{table}[th]
	\caption{The 50 models of InverseCSG. The error is the relative volumetric error: the volume of the symmetric
		difference between the result and the input, divided by the volume of the input. The values of InverseCSG are
		read from Fig.~8 of \citet{du2018inversecsg}, and its times are those it reports for sequential runs; our times
		are wall times on all 32 threads of a 16-core desktop CPU (AMD Ryzen 9 7950X).}
	\label{tab:icsg}
	\centering\small
	\begin{tabular}{@{}lrrrrrrr@{}}
		\toprule
		& \multicolumn{3}{c}{error, \%} & \multicolumn{2}{c}{models with} & \multicolumn{2}{c}{time per model, s} \\
		\cmidrule(lr){2-4} \cmidrule(lr){5-6} \cmidrule(l){7-8}
		method & mean & median & max & error $\le 1\%$ & lower error & mean & max \\
		\midrule
		InverseCSG & 0.89 & \textbf{0.40} & \textbf{6.6} & 37 & 22 & 2298 & 19818 \\
		Ours       & \textbf{0.83} & 0.46 & 10.4 & \textbf{43} & \textbf{28} & \textbf{21.5} & \textbf{202} \\
		\bottomrule
	\end{tabular}
\end{table}

\subsection{Ablation Study}
\label{sec:ablation}

To analyze importance of different parts of our pipeline we have split runs of \Cref{tab:main} into three groups. 
We have performed detailed analysis for a random subset of BenchCAD dataset (Table~\ref{tab:ablation}) and analyzed logs on full datasets (Table~\ref{tab:allstat}). 
In both cases, we observe a similar picture. In many cases (79\%--96\% of parts on DeepCAD, Fusion360, and BenchCAD) a hypothesis of the combiner, tuned once by the local search, already reaches $F^\ast$, and the genetic algorithm does not run a second generation. 
The exceptions are the mechanical-parts benchmarks MCB and CADENA-Bench, where evolutionary search runs past the first generation on more than two thirds of the parts. 
So we conclude that on sketch-and-extrude benchmarks the accuracy thus comes from detection and gradient descent, and the evolutionary search decides the result on mechanical parts. 
\Cref{fig:complexity} shows the same in terms of the size of the program: the error and the time grow with the number of solids, and programs of more than 16 solids are rare on DeepCAD and Fusion360 and make up nearly half of MCB (45\%) and half of CADENA-Bench (52\%).

\begin{table}[th]
	\caption{Detailed ablation on a fixed 1000-part subset of BenchCAD; one component is removed per row, with the seed of the full method. IoU is the mean, $\ge 0.99$ the share of parts with at least that IoU, $\Delta$ the paired difference in mean IoU to the full method, with a 95\% bootstrap interval over parts, and time the mean time per part in seconds on one thread.}
	\label{tab:ablation}
	\centering\small
	\begin{tabular}{@{}lrrlr@{}}
		\toprule
		configuration & IoU & $\ge 0.99$ & $\Delta$ [95\% CI] & time, s \\
		\midrule
		Full method                                & \textbf{0.997} & \textbf{97.0\%} & --                            & 221 \\
		\midrule
		Best hypothesis, no search                 & 0.987 & 80.4\% & $-0.010$ [$-0.013$, $-0.008$] & \textbf{17} \\
		Hypotheses + local search                  & 0.990 & 87.4\% & $-0.007$ [$-0.009$, $-0.005$] & \textbf{19} \\
		Search without hypotheses                  & 0.888 & 25.3\% & $-0.109$ [$-0.119$, $-0.099$] & 135 \\
		Search without local search                & 0.993 & 87.9\% & $-0.004$ [$-0.005$, $-0.003$] & 115 \\
		\midrule
		RANSAC-extrude detector only               & 0.973 & 81.1\% & $-0.024$ [$-0.030$, $-0.019$] & 162 \\
		RANSAC-direct detector only                & 0.993 & 93.5\% & $-0.004$ [$-0.006$, $-0.003$] & 222 \\
		Extrudes only (no revolves, no sweeps)     & 0.980 & 80.2\% & $-0.017$ [$-0.023$, $-0.013$] & 383 \\
		\bottomrule
	\end{tabular}
\end{table}

\begin{table}[th]
	\caption{Reconstruction statistics over all non-rejected targets of each dataset. Avg.\ solids and avg.\ params are the mean numbers of solid features and continuous parameters in the final program. Reached $F^\ast$ is the share of targets that reached the target fitness $F^\ast = 0.995$ before exhausting the 50{,}000-evaluation budget, and ``in 1st gen.'' is the share that reached it on the first generation (after primitive detection, combination, and one pass of gradient descent, without evolutionary offspring creation).}
	\label{tab:allstat}
	\centering\small
	\begin{tabular}{@{}lrrrr@{}}
		\toprule
		dataset & avg.\ solids & avg.\ params & reached $F^\ast$ & in 1st gen. \\
		\midrule
		DeepCAD      &  2.2 &  44 & 99.4\% & 96.3\% \\
		Fusion360    &  3.6 &  94 & 97.7\% & 88.8\% \\
		BenchCAD     &  6.6 & 283 & 93.5\% & 78.9\% \\
		MCB          & 18.3 & 772 & 62.2\% & 31.6\% \\
		CADENA-Bench & 20.6 & 983 & 59.8\% & 31.1\% \\
		\bottomrule
	\end{tabular}
\end{table}

\paragraph{Components.}
\Cref{tab:ablation} removes one component at a time, on a fixed subset of 1000 BenchCAD parts and with the seed of
the main runs. BenchCAD is the benchmark with the widest range of features, and the subset draws from all 106 of its
families in proportion to their size; on it, the full method has the mean IoU of the whole benchmark (0.997) and takes
221~s per part on average (245~s on the whole benchmark). Every ablation lowers the mean IoU, and no interval includes
zero. Over three seeds of the full method, the mean IoU changes by at most 0.0005 (0.9966--0.9971), and 1.2\% of the
parts change by more than 0.01, so every difference in the table is at least seven times the variation between seeds.

\emph{Hypotheses and search.} The best hypothesis of the combiner reaches a mean IoU of 0.987 on BenchCAD, and one
pass of local search over the initial population raises it to 0.990. Detection and local search thus take a part to
an IoU of 0.99 or more in 87\% of the cases, in 19~s on average, but leave 3.2 times the error of the full method.
The rest comes from the evolutionary search: on the 20\% of the parts where it runs past the first generation, it
raises the mean IoU from 0.957 to 0.990. Bevel gears, twisted brackets and twist drills account for two thirds of
this gain. The search without hypotheses starts from random programs, as earlier evolutionary methods do, and its
mutations draw no primitives from the pool. Its mean IoU is 0.888, and only 25\% of the parts reach 0.99, most of
them pins, keys, grommets and brackets made of a few extrudes or revolves; it runs 86 generations per part on
average against 10. Detection is thus what makes the evolutionary search work, and the search is what makes the
reconstruction of the harder parts precise.

\emph{Local search.} Without gradient descent, the genetic operators tune the continuous parameters by themselves.
The structure of the programs is mostly found (mean IoU 0.993), but they are less precise: 88\% of the parts reach
0.99 against 97\%, and the error is 2.4 times that of the full method, mostly on sweeps (twisted brackets, pipe
elbows) and gears. Each evaluation is cheaper without gradient descent, so the time per part is lower although the
search spends 2.6 times as many evaluations.

\emph{Detectors and vocabulary.} RANSAC-direct by itself loses much less than RANSAC-extrude by itself (mean IoU 0.993
against 0.973): it also detects revolves and sweeps, which the springs, twist drills, worm screws and tubes of
BenchCAD need, and these families hold two thirds of the error added by RANSAC-extrude alone. The two detectors
complement each other: with both, the error is 2.2 times lower than with RANSAC-direct alone, most of the difference
lies in duct elbows, handwheels and helical gears, and no family is reconstructed better by one detector alone.
Restricted to extrudes, the search approximates revolved and swept shapes by stacks of extrudes: the programs have
15 solids on average instead of 6, the time per part is 1.7 times higher, and the mean IoU drops to 0.980, mostly on
springs, bellows, worm screws and nozzles. The vocabulary of the program thus matters as much as the search over it.

\begin{table}[th]
	\caption{Other methods as primitive detectors. Top: CADFit on 300 Fusion360 parts. Bottom: the operations of CADFather's programs on 111 BenchCAD parts, selected as those where our result is the lowest (bevel and helical gears, mostly).}
	\label{tab:external}
	\centering\small
	\begin{tabular}{@{}llrr@{}}
		\toprule
		parts & configuration & mean IoU & median IoU \\
		\midrule
		Fusion360 (300)          & CADFit by itself                                & 0.969 & 0.9959 \\
		                         & Ours, RANSAC-direct + RANSAC-extrude            & \textbf{0.995} & \textbf{0.9996} \\
		                         & Ours, CADFit as the only detector               & 0.937 & 0.9974 \\
		                         & Ours, RANSAC-direct + CADFit                    & 0.976 & 0.9988 \\
		\midrule
		BenchCAD, hardest set (111) & CADFather by itself                             & 0.908 & 0.9012 \\
		                         & Ours                                            & 0.936 & 0.9367 \\
		                         & Ours + CADFather's operations as primitives     & \textbf{0.952} & \textbf{0.9492} \\
		\bottomrule
	\end{tabular}
\end{table}

\paragraph{Other methods as detectors.}
Primitive detection is the open part of the pipeline (\cref{sec:detection}): any method that outputs anything convertible to CSG trees can add them as primitives to the pool, and the combiner and the search treat them like the primitives of our own detectors. 
\Cref{tab:external} shows this for two methods. For CADFather, the extrudes and revolves of its final
program become primitives. We test this on about a hundred of the BenchCAD parts where our method performed worst, mostly bevel and helical gears, so the result is the possible gain, not an average one. With CADFather's primitives in the pool, the mean IoU rises from 0.936 to 0.952, that is 25\% reduction of volumetric error, at the same median time. The result is better that just using our method on more than half of the parts and never worse. It is also almost alvays above CADFather's own result: the search does not copy the program it was given, it combines its operations with our primitives and refines them.

CADFit proposes extrudes fitted to cross-sections of the target, and we run it in place of our RANSAC-extrude detector. 
On 300-model subset of Fusion360, the search seeded by RANSAC-direct and CADFit reaches a higher average IoU than CADFit by itself. CADFit's primitives are, however, weaker seeds than those of RANSAC-extrude, which gives 0.995, and CADFit takes most of the time of such a run (225~s on average against 3.3~s). 
A detector thus helps where it finds what the others miss, as CADFather's does on gears and creating specific detectors for complex, but common parts (gears, cases etc.).

\paragraph{CadQuery.}
\label{sec:exp_programs}

\begin{figure}[htbp]
	\centering
	\begin{minipage}[t]{0.34\linewidth}
		\vspace{0pt}\centering\textbf{CSG tree}
\begin{lstlisting}[basicstyle=\ttfamily\footnotesize]
Subtract
  Subtract
    Extrude(h=0.33)
      Profile P
    Extrude(h=1)
      Scale2D(0.7)
        Profile Q
  Translate3D(-0.05,0,0.75)
    Rotate3D(0,0,0.2)
      Extrude(h=0.5)
        Circle((0,0),0.25)
\end{lstlisting}
		\par\smallskip\textbf{Model}\par\smallskip
		\includegraphics[width=\linewidth,height=3cm,keepaspectratio]{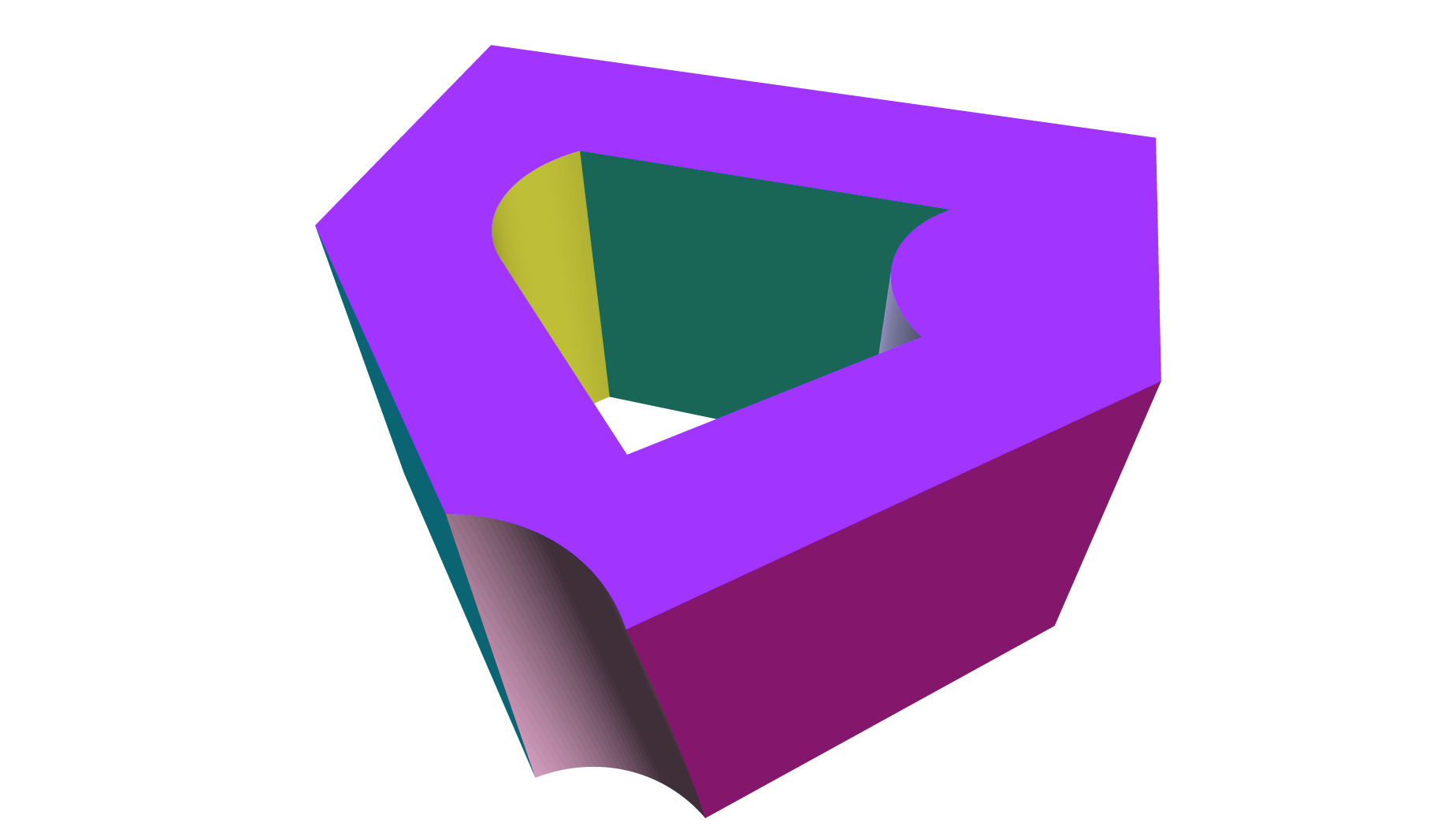}
	\end{minipage}\hfill
	\begin{minipage}[t]{0.62\linewidth}
		\vspace{0pt}\centering\textbf{CadQuery program}
\begin{lstlisting}[language=Python,basicstyle=\ttfamily\footnotesize]
import cadquery as cq
P = [(-0.7,-0.7), (0.7,-0.7), (0.7,0),
     (0,0.7), (-0.7,0)]
A = (cq.Workplane("XZ").polyline(P).close()
     .extrude(0.33, both=True))
Q = (cq.Workplane("XZ")
     .moveTo(-0.5,-0.5).lineTo(0.5,-0.5)
     .threePointArc((0.394949,-0.25), (0.5,0))
     .lineTo(0,0.5).lineTo(-0.5,0)
     .threePointArc((-0.605051,-0.25), (-0.5,-0.5))
     .close())
q = Q.val().scale(0.7)
B = (cq.Workplane("XZ").newObject([q]).toPending()
     .extrude(1, both=True))
C = cq.Workplane("XZ").circle(0.25)
C = C.extrude(0.5, both=True)
C = C.rotate((0,0,0), (0,0,1), 11.459156)
C = C.translate((-0.05,0,0.75))
result = A.cut(B).cut(C)
\end{lstlisting}
	\end{minipage}
	\caption{The demo model, shown using standard CadQuery calls. $P$ and $Q$ denote the
	profiles defined in the code. Extrusion parameters are half-heights; tree angles are in radians and CadQuery
	angles are in degrees. The $XZ$ workplane aligns the extrusions with the CSG frame. Values are rounded for display.}
	\label{fig:cadquery_demo}
\end{figure}

A reconstruction is useful only as a program that a CAD system executes and a designer can edit. That is why we provide a way to export our CSG trees to a CadQuery program. The exporter follows
the tree: it creates sketches, applies features and transformations, and combines the resulting solids with Boolean
operations. The dimensions come from the reconstructed tree. The resulting program can be executed independently
of ICAD, edited, and exported to STEP. \Cref{fig:cadquery_demo} illustrates this with demo model:
a polygonal extrusion with a scaled curved-profile cut and a tilted cylindrical cut.
The operation mappings, including the cases that require an approximation, are described in
\cref{sup:cadquery}.

\begin{figure*}[th]
	\centering
	\includegraphics[width=0.85\linewidth]{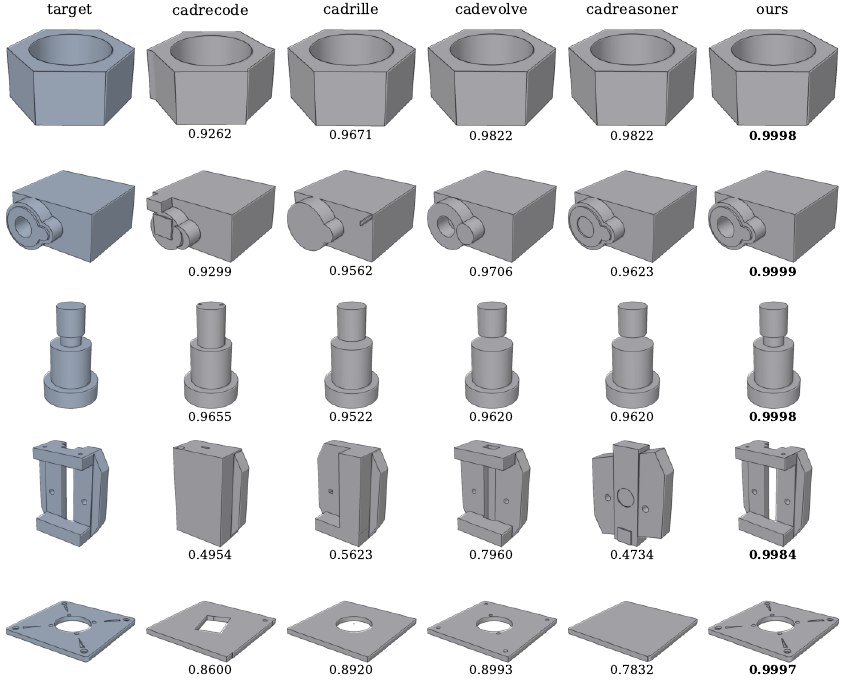}
	\caption{Qualitative comparison on Fusion360 parts, with the exact IoU against the target under each result.}
	\label{fig:qualitative}
\end{figure*}

\section{Conclusion and Future Work}
\label{sec:conclusion}
We present a novel algorithm for mesh-to-CAD reconstruction that extensively tested our approach on a range of benchmarks and showed that it consistently outperforms prior methods on different metrics. However, it has limitations, with lower quality and much higher optimization time on most complex models, so its integration with VLM-based approaches should be beneficial in these cases.

While our experiments show the superiority of proposed algorithm over the agent-based methods, we believe that the best results will be achieved by combining these two approaches what our preliminary experiments in this direction show (Table \ref{tab:external}). While often not able to solve the mesh-to-CAD task perfectly, VLMs are capable of higher level judgment that our purely algorithmic, non-learnable method lacks. Using one for primitive detection, guided mutations or at post-processing could improve the overall quality in hard cases albeit with some increase in reconstruction time and cost.

Another thing we would like to improve is handling of larger and more complex models. Current algorithm is targeted at reconstructing what is generally a one detail or a simple, few-parts assembly. Moving to complexity typical for CAD assemblies (hundreds of primitives, many thousands parameters) would require some task decomposition methods, parallel optimization of parts or other ways to tackle increasing complexity.

\bibliographystyle{unsrtnat}
\bibliography{references} 

\pagebreak

\appendix

\section{Kernel Details}
\label{sup:kernel}

\subsection{Fields}
\label{sec:kernel_fields}

Each node $N$ defines a field $f_N$, on $\mathbb{R}^3$ for a solid and on $\mathbb{R}^2$ for a sketch, whose negative
set is its solid, $S(N) = \{p : f_N(p) < 0\}$. An inner node computes its field from those of its children.
Booleans, in 2D and in 3D alike, are the usual~\citep{ricci1973constructive}
\begin{equation}
	f_{A \cup B} = \min(f_A, f_B), \qquad f_{A \cap B} = \max(f_A, f_B), \qquad f_{A \setminus B} = \max(f_A, -f_B).
	\label{eq:booleans}
\end{equation}
A transform maps the query point into the frame of its child, $f_{\mathcal{R}(A)}(p) = f_A(\mathcal{R}^{-1}p)$ for a
rigid motion $\mathcal{R}$, and a scale by $s$ gives $s\,f_A(p / s)$.

\begin{figure}[th]
	\centering
	\includegraphics{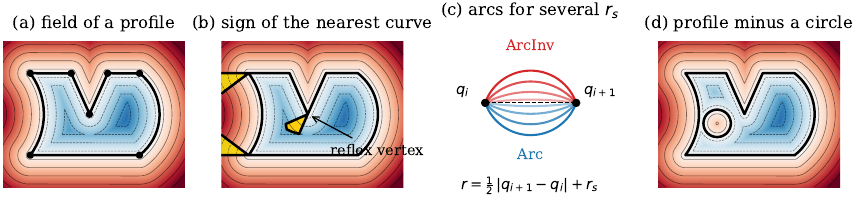}
	\caption{Sketch fields. A loop is a closed chain of lines and arcs with shared end points, so every parameter value
		is a closed loop; its sign comes from ray-crossing parity.}
	\label{fig:kernel_sketch}
\end{figure}

\paragraph{Loops.}
The field of a loop is its signed distance (\cref{fig:kernel_sketch}(a)). Its magnitude is the distance to the nearest of its curves, each a segment or an arc with a closed-form distance. 
To calculate the sign we cast ray from the query point along x axis, intersect it with curves and track parity. It is more robust near sharp angles than just checking on which side of 
the closest curve we are (\cref{fig:kernel_sketch}(b)). A circle has $f(q) = \lVert q - c \rVert - r$ (\cref{fig:kernel_sketch}(c)). Different profiles can be combined with boolean 
operations in 2D (\cref{fig:kernel_sketch}(d)).

\paragraph{Extrude and revolve.}
With $f_S$ the field of the sketch, an extrude has the standard field of an extrusion~\citep{quilez_sdf},
\begin{equation}
	f_{\mathrm{ext}}(p) = \min\bigl(\max(w_1, w_2), 0\bigr) + \bigl\lVert \max(w, 0) \bigr\rVert,
	\qquad w = \bigl(f_S(p_x, p_z),\; |p_y| - h\bigr),
\end{equation}
the signed distance to the negative quadrant in the plane of $(f_S, |p_y| - h)$, and a revolve evaluates its sketch
in the half-plane through the point and the axis,
\begin{equation}
	f_{\mathrm{rev}}(p) = f_S\Bigl(\sqrt{p_x^2 + p_z^2},\; p_y\Bigr).
\end{equation}
Both are exact signed distances when $f_S$ is.

\paragraph{Sweep.}
A sweep projects the query point $p$ to the closest point $\gamma^\ast$ of its path, expresses the offset
$p - \gamma^\ast$ in the frame $(u, v)$ there, and evaluates its sketch at it:
\begin{equation}
	g(p) = f_S\bigl(u \cdot (p - \gamma^\ast),\; v \cdot (p - \gamma^\ast)\bigr).
\end{equation}
A closed sweep takes $g$ as its field. An open sweep must also stop at its two ends. Where $\gamma^\ast$ is an end of
the path, the end cap joins $g$ the way an extrude joins its side and caps, with $|p_y| - h$ replaced by the offset
of $p$ past the end plane along the end tangent. Elsewhere the field is
$\max\bigl(g, -\min(\lVert \gamma^\ast - k_1 \rVert, \lVert k_m - \gamma^\ast \rVert)\bigr)$, which keeps it
continuous with the caps. Where two straight spans meet, the frame pivots from one direction to the other, which
rounds the corner the way a ball joint of the sketch would. Turns of a spring can touch or fuse, as the ground end
coils of a compression spring do. There the closest point of the path may be an end while $p$ lies inside the
neighbouring turn, so an open sweep evaluates its sketch a second time, at the closest perpendicular foot of $p$ on
the path that is not an end, and takes the smaller of the two values.

\paragraph{Loft.}
A loft is the ruled solid between two sketches, $S_0$ in the plane $y = -h$ and $S_1$ in the plane $y = h$. The two
sketches must be compatible: the same tree of 2D Booleans, whose paired loops are either two circles or two profiles
with as many points and the same curve type at every span. Point $a_i$ of a loop of $S_0$ is paired with point $b_i$ of
its loop in $S_1$, and a loop may collapse to a point, the apex of a cone or a pyramid. At the height of the query
point, $\tau = \operatorname{clamp}\bigl((p_y + h) / 2h,\, 0,\, 1\bigr)$, the section $S(\tau)$ is the same sketch tree
with every point at
\begin{equation}
	q_i(\tau) = (1 - \tau)\, a_i + \tau\, b_i,
\end{equation}
and with the center and the radius of every circle interpolated in the same way. An arc runs between its interpolated
ends and spans the angle interpolated between the angles of its two arcs. A line span therefore traces a bilinear
patch, and a pair of arcs of equal angle traces the ruled surface between them, since every point of an arc of a given
angle is linear in its ends. This is the ruled loft of OCCT, which CadQuery~\citep{cadquery} builds from two
sections. OCCT may pair the vertices of a twisted loft differently, so for export the loft is passed as intermediate
sections at several values of $\tau$, which keep its pairing. The field is the field of the section, divided by a
Lipschitz bound and capped by the two section planes:
\begin{equation}
	f_{\mathrm{loft}}(p) = \max\Bigl(\frac{f_{S(\tau)}(p_x, p_z)}{L},\; |p_y| - h\Bigr),
	\qquad L = \sqrt{1 + \Bigl(\frac{v}{2h}\Bigr)^2},
\end{equation}
where $v$ bounds how far any point of the section moves per unit of $\tau$: the largest $\lVert b_i - a_i \rVert$ of
a pair of profiles, $\lVert c_1 - c_0 \rVert + |r_1 - r_0|$ of a pair of circles, and a bound from the ends and the
angles of a pair of arcs. The section field is 1-Lipschitz in $(p_x, p_z)$ and changes by at most $v / 2h$ per unit of
height, so it is $L$-Lipschitz in 3D. The quotient is therefore 1-Lipschitz and never overestimates the distance
(\cref{sec:kernel_pseudo}). Near the sides it lies between $1/L$ of the distance and the distance itself, and it is
exact on a right circular cone. With $S_1 = S_0$ the loft has the solid of an extrude, with $v = 0$ and $L = 1$.

\subsection{Pseudo-SDFs and Truncated Evaluation}
\label{sec:kernel_pseudo}

The search relies on four properties of a field: its sign, which decides occupancy and therefore IoU; its zero set,
which is the surface of the exported model; its values near the surface, which gradient descent compares with the
signed distance to the target (\cref{sec:gd}); and the bound $|f(p)| \le \operatorname{dist}(p, \partial S)$, which
lets the IoU estimator decide a whole block of space from the value at its center (\cref{sec:fitness}). An exact SDF
has all four. The Booleans of \cref{eq:booleans} keep the sign and the zero set; their result is not an exact
distance, but it is 1-Lipschitz whenever the children are and so never overestimates the
distance~\citep{hart1996sphere}. The extrude and revolve formulas above preserve this property too, so any tree of
extrudes and revolves has a 1-Lipschitz field that is exact near its surface, except close to the edges where
Booleans meet. The sweep field is exact near its surface, where the closest point of the path is unique and the
offset lies in its normal plane. Farther away it keeps the sign and the zero set, but nothing bounds its slope, and
neither the exact distance to a swept solid nor its truncated version is cheap to compute. We call fields of this
kind, correct in sign and zero set and exact only near the surface, pseudo-SDFs.

Away from the surface the exact value is rarely needed. The kernel therefore also evaluates a truncated field with a
threshold $\tau$: it returns $f(p)$ whenever $|f(p)| < \tau$, and otherwise any value that has the sign of $f(p)$ and
a magnitude above $\tau$. The threshold is propagated down the tree, where a scale divides it and a blend widens it by
the width over which it reads its children, and the expensive nodes use it to stop early: the search for the nearest
curve of a profile ignores curves farther than $\tau$, and a sweep whose path is farther from the point than the
radius of its sketch plus $\tau$ is not evaluated at all. The IoU estimator asks for $\tau$ equal to the half-diagonal
of a block, and the occupancy signatures of \cref{sec:ga} ask for $\tau = 0$, that is, for the sign only.

\subsection{Fitness}
\label{sec:fitness}

Space is divided into a hierarchy of blocks. A block is
decided for the target when the voxel pyramid shows it entirely inside or entirely outside, and for the program when
the field at its center exceeds the block's half-diagonal in magnitude (\cref{sec:kernel_pseudo}). Blocks decided for
both shapes add their volume to the intersection and the union exactly; only blocks on either surface are split into
the next level, and at the finest level they are sampled by Monte Carlo, one point per block in each round, until a
confidence interval on the IoU (a stratified ratio interval by the delta method) is narrower than
$0.1\,(1 - \widehat{\mathrm{IoU}})$, or the sample budget is spent. The precision is therefore relative to the
remaining error: a candidate near 0.5 is ranked to within $\pm 0.05$ and one near 0.99 to within $\pm 0.001$, which
is what selection needs. The search uses two settings: a coarse one ($16^3$ finest blocks, at most $5 \cdot 10^4$
samples, interval $0.33\,(1 - \widehat{\mathrm{IoU}})$) to screen every program, and a precise one ($32^3$ blocks, at
most $10^6$ samples) for the programs that pass the screening. All programs are scored on the same sample points, so
their differences are not blurred by independent sampling noise, and the occupancy of the target at those points is
cached between evaluations. The fitness also subtracts the volume of the program's bounding box that lies outside the
scene, which keeps material from escaping where no sample can see it.

\begin{figure}[th]
	\centering
	\includegraphics[width=0.85\linewidth]{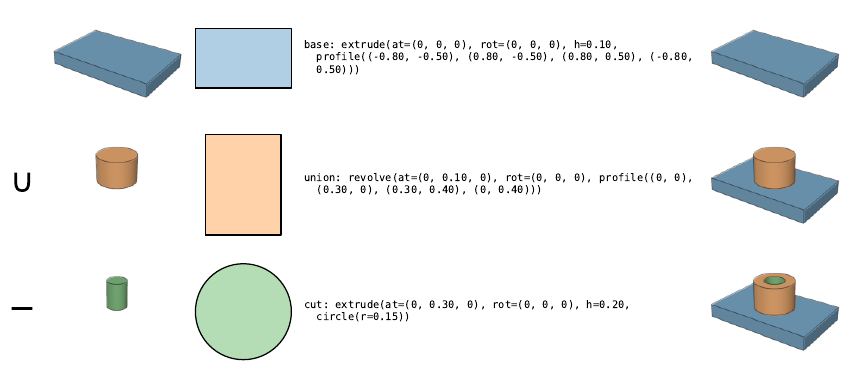}\\[2mm]
	\includegraphics[width=\linewidth]{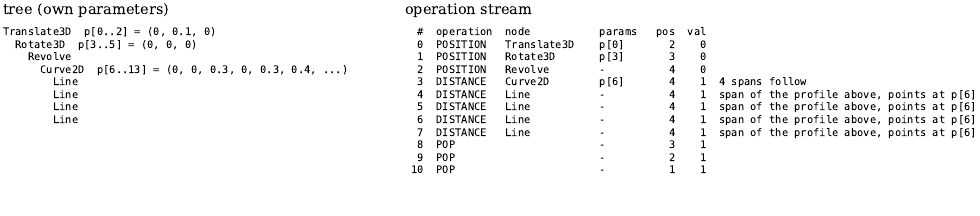}
	\caption{Top: a program of three steps, each gene with its sketch, its text and the solid built so far. Bottom: the second gene (the boss), its tree with its own parameters, and its stream of operations with the depths of the position and value stacks.}
	\label{fig:opstream}
\end{figure}

\subsection{Operation stream.}
A tree is compiled by a depth-first walk. A transform node emits a position operation, the stream of its child, and a
distance operation that pops the position; a feature emits a position operation that projects the point into its
sketch's plane, the stream of its sketch, and a distance operation that joins the sketch's value with its own terms;
a Boolean emits the streams of its two children and a distance operation that combines the two top values of the
stack. A profile is one operation followed by one sub-operation per span, which carries the span's type and the
offset of its first point in the parameter array, so that consecutive spans share their points. Each value on the
stack carries, besides the distance, the parameter offset of the leaf that determined it, which labels every face of
a result's mesh with its primitive. \Cref{fig:opstream} shows a program and the stream of one of its genes.

\section{RANSAC-extrude Primitive Detector}
\label{sup:extrude_combiner}

RANSAC-extrude runs the shape engine of \cref{sec:detection} with planes and cylinders only (band $\epsilon = 10^{-3}$
of the largest side of the bounding box, normals within $\arccos 0.95 \approx 18^\circ$, minimal area $10^{-4}$ of
the surface, at most 5000 candidates), on a copy of the mesh that is welded and refined by longest-edge
bisection~\citep{rivara1997lepp}, so that every shape owns whole, small triangles. It then builds three kinds of
extrudes from the shapes. The conflict keys of a primitive are the ids of the shapes it was built from, so two
extrudes built from one plane are rivals.

\paragraph{Cap pairs.}
Planes are grouped by normal direction, and the planes of a group are split by their offset along the normal into
tops and bottoms, which face in opposite directions. Every combination of top and bottom planes of a group, up to 64
per group, is a candidate extrude: its extent along the normal is that of its planes, and its profile is the outline
of their triangles, rasterized on a bitmap fitted to their bounding box ($1024^2$ pixels), traced into contours,
cleaned of contours smaller than 16 pixels and simplified by the algorithm of \citet{visvalingam2017line}. Several
contours make one sketch with several loops. Every triangle votes, by its area, whether its normal points away from
the extrude's center (a solid) or towards it (a cut); 80\% of the area decides, and an extrude with no clear majority is
dropped. A group with planes on one side only is either a thin solid whose two faces the RANSAC merged into one plane,
when the votes are within a factor of five of each other, or the floor of a cut whose depth the mesh does not show. Such
a cut is extruded outwards from the floor, about as deep as the longer side of its profile and 10\% wider than its
outline, so that it reaches past the part.

\paragraph{Cylinders.}
A detected cylinder spans the extent of its triangles along its axis. If its triangles face outwards over one range
of the axis and inwards over another, as on a cup with a floor, it is split into a boss and a hole with conflict keys
of their own. A cylinder that covers enough of a turn becomes an extrude of a circle; narrower arcs are left to the
side groups below.

\paragraph{Side groups.}
Candidate extrusion axes are taken from the directions to which many shapes are orthogonal: the normals of planes and
the axes of cylinders, discretized. Noisy normals give many near-duplicate axes, so a group of shapes is dropped when a
larger group already explains 90\% of its shapes, and at most 16 groups are kept. The shapes of a group are projected
onto a plane orthogonal to its axis, a plane becoming a segment and a cylinder a polyline of its arc, and segments
whose ends are within 0.03 of each other are linked into chains. Every closed chain is the profile of an extrude
whose height and polarity come from its triangles.

\section{RANSAC for Common CSG Primitives}
\label{sup:detectors}

RANSAC-direct runs the same engine with plane, extrude and revolve shapes, and sweeps when asked for, and turns every
shape into one primitive or two.
The mesh is welded, cleaned of duplicate triangles, consistently oriented and refined by longest-edge bisection until
no edge is longer than 5\% of the largest side of its bounding box, and it is scaled so that this side has a fixed
length; all tolerances are relative to it. Extraction runs in rounds. In each round, candidates of a type drawn
uniformly at random are generated and scored by their support, the area of the triangles they explain, until the
chance that the best candidate of the round, or any shape of the minimal area, has not yet been drawn falls below
$10^{-3}$. The best candidate is extracted, its triangles are retired, and the rounds continue until less than the
minimal area is left, or 128 shapes or $10^5$ candidates are reached. A plane orthogonal to the axis of an extrude it
touches is a piece of that extrude's side, such as a wall cut shorter than the rest, and joins the extrude instead of
becoming a shape of its own.

\paragraph{Extrudes.}
A Gauss map search of the whole mesh, seeded with the coordinate axes, returns up to eight axes to which much of the
surface is orthogonal and which stand out against their neighbourhood by a factor of 1.5. A candidate draws one of them,
weighted by the live area orthogonal to it, and a seed triangle on that area. Its side grows from the seed over
connected triangles whose normals are orthogonal to the axis to within $5^\circ$, and the axis is refitted to the side
as the smallest eigenvector of $\sum_i w_i n_i n_i^\top$, reweighted so that triangles the axis does not explain lose
their weight. Every side triangle, seen along the axis, is a short segment with an outward normal and a range of
heights; the segments are deposited into an evidence grid in the base plane, and the profile is traced from the grid as
a closed loop and simplified to within $2 \cdot 10^{-3}$ of the bounding box. Side triangles far from the loop, and
walls much higher or lower than the seed's wall, are left for other candidates: they belong to a cut inside the profile
or to another extrude standing next to this one. Caps are the triangles that face along the axis, lie in the base or
the top plane and project inside the loop. A thin solid, such as a rib, is two walls back to back that the grid can
merge into an open loop; such evidence is deposited again into a narrow strip of finer cells. The score of a candidate
is the area of its side and caps.

\paragraph{Revolves.}
On a surface of revolution, every normal line meets the axis. A line with direction $a$ through a point $c$ meets the
normal line $(p, n)$ of a triangle if $a \cdot (p \times n) + n \cdot (c \times a) = 0$, which is linear in the
Plücker coordinates $(a, \bar a = c \times a)$ of the axis. The axis of a patch around the seed therefore minimizes
$\sum_i w_i \bigl(a \cdot (p_i \times n_i) + n_i \cdot \bar a\bigr)^2$ subject to $\lVert a \rVert = 1$; eliminating
$\bar a$ leaves $a$ as the smallest eigenvector of a $3 \times 3$ Schur complement, and the fit is reweighted a few
times. Patches that are flat, pieces of a sphere, or whose normal lines share no axis are rejected. The members of a
revolve are the triangles whose normals lie in the plane through them and the axis, to within $5^\circ$. Seen in the
$(u, v)$ half-plane of the axis, they are deposited into an evidence grid like the side of an extrude, cells that hold
less than 5\% of a full turn are emptied, which removes flat faces that cross the revolve over a narrow sector only, and
the profile is traced from the grid. A profile that is open, such as the outside of a solid of revolution, is closed
to the axis.

\paragraph{Sweeps.}
A sweep is traced in the way the centerline of a vessel or a pipe is traced in a scan~\citep{tagliasacchi2009curve}.
From a seed patch, which gives the first tangent and center, the surface is cut by a short slab orthogonal to the
local direction of the path, and the cross-section in the slab is fitted. The tangent is the direction the slab's
normals are orthogonal to. A circular cross-section is fitted as the envelope of the slab's tangent lines,
$n_i \cdot (p_i - c) = r$, which is linear in $(c, r)$, does not penalize the flat facets of a coarse tessellation, and
rejects a fused neighbouring coil as outliers; normals that point at the path make a bore, a hole swept along it. Any
other cross-section is a polygon around the point closest to the slab's normal lines. The slab then steps half a
cross-section along the path in both directions, until it is empty, its cross-section changes size by 30\%, explains
less than 30\% of its slab or shows less than half a turn, or the path closes on itself. A thread, of which only part
of the cross-section shows above the solid it stands on, is traced again as an open cross-section when a closed one
fails. The knots of the path are placed every $22.5^\circ$ of turning, so a helix has 16 knots per turn, and the frame
axis is fitted to the path. Because the vectorized evaluator handles straight spans only (\cref{sec:kernel_eval}), the
path is emitted as a polyline.

\paragraph{Polarity and primitives.}
For an extrude, the side triangles of a solid face out of the profile's loop and its caps face out of the prism. Let
$A^+$ and $A^-$ be the areas that agree and disagree with this; the vote $(A^+ - A^-)/(A^+ + A^-)$ makes a solid if it
is above $-0.8$ and a cut if it is below $0.8$, and both, with a shared conflict key, in between. Both are also emitted
when the sides and the caps each carry at least a tenth of the area but vote in opposite directions, as on a boss
whose top is the floor of a pocket. A profile that fits a circle becomes a circle. A cut keeps an end that has caps,
which is a floor, and moves an open end outwards by the larger of $2\epsilon$ and 5\% of its height, since an open end
is where the cut breaks out into the air. Revolves are treated in the same way along their axis. A plane that joined
no extrude becomes a cut: the footprint of its triangles, rasterized on a $512^2$ bitmap, traced and simplified, swept
from the plane outwards as far as the footprint is wide. A sweep becomes a solid, or a cut if it is a bore.

\section{Mutation Operators}
\label{sup:mutations}

\Cref{tab:mutations} lists the mutation operators and their default weights. An operator that has nothing to do in a
program, such as \texttt{jump\_axis} in a program without revolves, returns without a change, and another operator is
drawn, up to 32 times. The actual frequencies of the operators therefore depend on the programs.

\begin{table}[ht]
	\caption{Mutation operators. $p$ is the probability that an operator is drawn first, from the weights; operators
	with weight 0 (\texttt{wrap\_steps}, \texttt{unwrap}, \texttt{move\_into\_subtree}, which create and remove nesting)
	are left out.}
	\label{tab:mutations}
	\centering\small
	\begin{tabularx}{\linewidth}{@{}llrrX@{}}
		\toprule
		group & operator & weight & $p$, \% & effect \\
		\midrule
		structure & \texttt{add\_step}     & 1    & 3.5  & inserts a step anywhere: a pool primitive (prob.\ 0.5) with the op of its polarity, or a random gene inside what the spine builds, cut with prob.\ 0.5 \\
		          & \texttt{remove\_step}  & 1    & 3.5  & removes a step; the base only if the next step is not a cut \\
		          & \texttt{move\_step}    & 1    & 3.5  & moves a step to another position of its spine; the base only past a union \\
		          & \texttt{flip\_op}      & 0.5  & 1.8  & union $\leftrightarrow$ cut \\
		          & \texttt{pool\_swap}    & 0.5  & 1.8  & replaces a pool gene by a pool primitive that shares a conflict key, with the op of its polarity \\
		frame     & \texttt{jump\_position} & 0.5 & 1.8  & moves a gene by up to 0.5 along each axis \\
		          & \texttt{jump\_angles}  & 0.5  & 1.8  & turns a gene by up to $90^\circ$ about each axis; a locked one to random right angles \\
		          & \texttt{toggle\_axis\_lock} & 0.25 & 0.9 & locks a frame (snapping it to the nearest axis-aligned one) or unlocks it \\
		sketch    & \texttt{add\_point}    & 10   & 35.4 & splits a line of a profile at its midpoint \\
		          & \texttt{remove\_point} & 10   & 35.4 & removes a point of a profile with more than three \\
		          & \texttt{add\_hole}     & 0.5  & 1.8  & cuts a random loop out of a sketch around the centroid of its base loop \\
		          & \texttt{remove\_hole}  & 0.5  & 1.8  & removes a hole loop \\
		          & \texttt{circle\_to\_polygon} & 0.25 & 0.9 & a circle becomes a regular polygon of 3 to 12 points and the same area \\
		          & \texttt{polygon\_to\_circle} & 0.25 & 0.9 & a profile becomes the circle of its area at its centroid \\
		extrude   & \texttt{jump\_height}  & 0.5  & 1.8  & multiplies the half-height by a factor in $[0.5, 1.5]$ \\
		revolve   & \texttt{jump\_axis}    & 0.5  & 1.8  & tilts the axis about its origin; a locked one by a right angle \\
		          & \texttt{extrude\_to\_revolve} & 0.25 & 0.9 & an extrude of concentric circles becomes the revolve of rectangles $[0, r] \times [-h, h]$ \\
		          & \texttt{revolve\_to\_extrude} & 0.25 & 0.9 & the inverse, for rectangles with one $v$ range; rings where a rectangle leaves the axis \\
		\bottomrule
	\end{tabularx}
\end{table}

New random genes made by \texttt{add\_step} are extrudes with probability 0.8 and revolves with probability 0.2. An extrude has a half-height in $[0.02, 0.25]$ and a sketch that is a circle (probability 0.3) or a
star-shaped blob of 3 to 12 points with radii in $[0.05, 0.6]$. A revolve has a rectangle (a cylinder or a tube) or a
lathe profile whose outer radius varies along the axis; half of them reach the axis, the others are tubes. Half of all
new genes have locked frames. Sweep genes come only from the detectors: no operator creates one, but all operators
that act on frames and sketches act on them, and gradient descent moves their knots.

\section{CadQuery Mapping}
\label{sup:cadquery}

\Cref{tab:cadquery} gives CadQuery calls for direct mappings and pseudocode for composite constructions.
Here \texttt{W} denotes \lstinline|cq.Workplane("XY")|, $A,B$ are built objects, and \texttt{w} is a sketch wire.
Sketch coordinates $(u,v)$ become $(x,z)$ on extrusion, with $Y\in[-h,h]$. Angles are converted to degrees.

\begin{table}[ht]
	\caption{CSG-to-CadQuery mapping. Composite constructions are written as pseudocode; field contours and section
	lofts provide approximations where a direct mapping is unavailable.}
	\label{tab:cadquery}
	\centering\footnotesize
	\begin{tabularx}{\linewidth}{@{}p{0.30\linewidth}X@{}}
		\toprule
		program & CadQuery / pseudocode \\
		\midrule
		gene frame: Translate3D, Rotate3D & \lstinline|A.translate(t)|; rotations about $X$, then $Y$, then $Z$,
		e.g.\ \lstinline|A.rotate((0,0,0), (1,0,0), ax)|. \\
		\midrule
		Circle($c$, $r$) & \lstinline|W.moveTo(cx, cy).circle(r)|. \\
		Box2D($c$, $a$, $b$) & \lstinline|W.moveTo(cx, cy).rect(2*a, 2*b)|; $a,b$ are half-sizes. \\
		Profile with Line spans & \lstinline|W.polyline(points).close()|. \\
		Arc / ArcInv span & \lstinline|.threePointArc(mid, end)|; compute the arc midpoint from its endpoints,
		radius parameter and side. \\
		\midrule
		Sketch $A\setminus B$, $A\cup B$, $A\cap B$, $A\mathbin{\triangle}B$ & Apply the extrude/revolve/sweep to each
		profile, then combine: \lstinline|result <- feature(A) op feature(B)|. Standalone sketches: trace the region boundary. \\
		\midrule
		Translate2D, Rotate2D & \lstinline|w.translate((tx,ty,0))|;
		\lstinline|w.rotate((0,0,0),(0,0,1),a)|. \\
		Scale2D, Scale3D & \lstinline|shape.scale(k)|, about the origin. \\
		Mirror2D in $X$ or $Y$ & \lstinline|w.mirror("XZ")| or \lstinline|w.mirror("YZ")|. \\
		Mirror3D in $XY$, $XZ$, $YZ$ & \lstinline|A.mirror("XY", union=False)|, likewise for $XZ$ and $YZ$. \\
		Extrude($h$) &
		\lstinline|w.extrude(h, both=True).rotate((0,0,0),(1,0,0),90)|. \\
		Revolve($\theta$) & Clip negative radii, then
		\lstinline|w.revolve(a, (0,0,0), (0,1,0))|; a full turn uses $a=360$. \\
		\midrule
		3D path: Line, Spline spans & \lstinline|cq.Edge.makeLine(p0, p1)|;
		Catmull--Rom spans become \lstinline|cq.Edge.makeBezier(controls)|. \\
		Sweep, SweepClosed & For one straight span with a centered circle or a constant profile frame:\newline
		\lstinline|placed_profile.sweep(path, combine=False)|.\newline
		Otherwise, \lstinline|support <- ribbon from path to its copy (2x,y,2z)|;\newline
		\lstinline|place profile; follow path oriented by support|;\newline
		\lstinline|round joints; cap ends or close seam; make solid|. \\
		\midrule
		Loft($h$, $S_0$, $S_1$) & \lstinline|sections <- corresponding profiles from Y=-h to Y=h|;\newline
		\lstinline|W.newObject(sections).loft(ruled=True)|. Insert sections when pairing or arc angles change;
		a collapsed end becomes an apex. \\
		\midrule
		Sphere, Box3D, Cylinder & \lstinline|W.sphere(r)|, \lstinline|W.box(2*a,2*b,2*c)|, or
		\lstinline|W.circle(r).extrude(h,both=True)|; translate to the center. Cylinder axis: $Z$. \\
		steps $\cup$, $\setminus$, $\cap$ & \lstinline|A.union(B)|, \lstinline|A.cut(B)|,
		\lstinline|A.intersect(B)|. \\
		XOR ($A\mathbin{\triangle}B$) & \lstinline|A.cut(B).union(B.cut(A))|. \\
		ExtrudeChamfer, ExtrudeFillet & Treat only the extrusion rims:
		\lstinline|s.faces(">Y or <Y").edges().chamfer(d)| or \mbox{\lstinline|.fillet(d)|}.
		Other profile/radius cases use sampled sections. \\
		\midrule
		Union / Intersect / Subtract with Chamfer or Fillet & \lstinline|f <- modified Boolean field(A, B, d)|;\newline
		\lstinline|result <- trace contours; loft if solid|. \\
		SmoothUnion($k$) & \lstinline|f <- quadratic smooth minimum(A.field, B.field, k)|;\newline
		\lstinline|result <- loft contours of f|. \\
		\midrule
		Offset($d$), sketch or solid & Positive $d$ shrinks and negative $d$ expands. Adjust primitive dimensions,
		e.g.\ \lstinline|W.circle(r-d)|, or \lstinline|trace/loft contours of (base.field + d)|. \\
		\midrule
		Bounded complements & $A\cap\operatorname{Inverse}(B)$ becomes \lstinline|A.cut(B)|; two inversions cancel. \\
		Ring, ExtrudeStack & \lstinline|ring <- outer circle minus inner circle|;\newline
		\lstinline|stack <- union of placed extrusions|. \\
		\bottomrule
	\end{tabularx}
\end{table}

Traced contours retain holes and disconnected regions. Other sweeps use an auxiliary support surface to control the frame; difficult
joints may use pipe pieces or section lofts. Spatial sweep corners, changing-arc lofts and sampled fields can require
approximations. Standalone unbounded complements are unsupported. CAD validity and shell closure are checked on
execution.

\section{Parameters}
\label{sup:params}

\Cref{tab:params} lists the settings used for every benchmark. The learning-rate multiplier of the local search is
\begin{equation}
	m(F) = \begin{cases}
		2 - F / 0.75 & F < 0.75, \\
		1 - 0.9 \bigl((F - 0.75) / 0.25\bigr)^2 & F \ge 0.75.
	\end{cases}
\end{equation}

\begin{table}[ht]
	\caption{Settings of the method (config/v7\_3.blk). Lengths are in scene units, where the part fits into $[-1, 1]^3$.}
	\label{tab:params}
	\centering\small
	\begin{tabularx}{\linewidth}{@{}lXr@{}}
		\toprule
		stage & setting & value \\
		\midrule
		target       & voxel grid, repair and fitness                         & $512^3$ \\
		             & near-surface samples for gradient descent; their band  & $2.5 \cdot 10^5$; 0.025 \\
		             & fixed volume samples for the pool                      & $2.5 \cdot 10^5$ \\
		detectors    & band half-width $\epsilon$, relative to the largest side & $10^{-3}$ \\
		             & minimal shape area, share of the surface               & $10^{-4}$ \\
		             & RANSAC-extrude: normal threshold; candidates; bitmap   & 0.95; 5000; $1024^2$ \\
		             & RANSAC-direct: revolves, sweeps                        & on, on \\
		combiner     & volume cover that makes two primitives rivals          & 0.9 \\
		             & maximal sets per component; hypotheses                 & $10^4$; 2500 \\
		GA           & population $N$; elites $E$; tournament                 & 512; 16; 16 \\
		             & crossover; random offspring                            & 0.5; 0.05 \\
		             & evaluation budget; target fitness $F^\ast$             & $5 \cdot 10^4$; 0.995 \\
		             & reset after generations without a gain of              & 50; 0.01 \\
		             & look-alike signature points                            & 4096 \\
		fitness      & coarse: finest blocks, samples, relative interval      & $16^3$, $5 \cdot 10^4$, 0.33 \\
		             & precise: finest blocks, samples, relative interval     & $32^3$, $10^6$, 0.1 \\
		             & confidence level $z$                                   & 1.96 \\
		local search & Adam steps; batch; base learning rate                  & 75; 256; $2 \cdot 10^{-4}$ \\
		post-process & points, uniform in the target's box; the box's margin  & $2.5 \cdot 10^5$; 0.05 \\
		             & IoU loss per edit; final acceptance                    & $10^{-3}$; $5 \cdot 10^{-4}$ \\
		             & profile deviation; circle fit; attempts                & 0.005; 0.002; 3 \\
		\bottomrule
	\end{tabularx}
\end{table}

\end{document}